\documentclass[epj]{svjour}

\usepackage{graphicx}
\usepackage{amsmath}
\usepackage{amssymb}
\usepackage{bm}

\begin{document}

\title{Modeling DNA beacons at the mesoscopic scale}

\author{Jalal Errami\inst{1} \and Michel Peyrard\inst{1}
\and 
Nikos Theodorakopoulos\inst{2}$^{,}$\inst{3}}

\institute{Laboratoire de Physique, ENS-Lyon, 46 all\'ee d'Italie,
        69364 Lyon Cedex 07, France 
\and
Theoretical and Physical Chemistry Institute,
National Hellenic Research Foundation, \\
Vas. Constantinou 48, 11635 Athens, Greece
\and
Fachbereich Physik der Universit\"at Konstanz,
Fach M 686, 78457 Konstanz, Germany}

\authorrunning{J. Errami, M. Peyrard, N. Theodorakopoulos}

\date{\today}

\abstract{We report model calculations on DNA single strands which
describe the equilibrium dynamics and kinetics of hairpin formation
and melting. Modeling is at the level of single bases. Strand rigidity
is described in terms of simple polymer models; alternative
calculations performed using the freely rotating chain and the
discrete Kratky-Porod models are reported. Stem formation is modeled
according to the Peyrard-Bishop-Dauxois Hamiltonian. The kinetics of
opening and closing is described in terms of a diffusion-controlled
motion in an effective free energy landscape. Melting profiles,
dependence of melting temperature on loop length, and kinetic time
scales are in semiquantitative agreement with experimental data
obtained from fluorescent DNA beacons forming poly(T) loops. Variation
in strand rigidity is not sufficient to account for the large
activation enthalpy of closing and the strong loop length dependence
observed in hairpins forming poly(A) loops. Implications for modeling
single strands of DNA or RNA are discussed.
} 

\maketitle

\section{Introduction}
\label{sec:intro}

DNA beacons are made of short single strands of DNA with terminal
regions consisting of complementary base sequences.  As a result the
two end-regions can self-assemble in a short DNA double helix, called
the stem, while the remaining central part of the strand forms a
loop. In this closed configuration, the single strand has the shape of
a hairpin. Such hairpin conformations are present in the secondary
structure of long single strands of DNA or RNA. A short single strand
of DNA which can form a hairpin becomes a so-called ``DNA beacon''
when one of its ends is attached to a fluorophore while the second end
is attached to a quencher. When the fluorophore and the quencher are
within a few Angstr\"oms of each other, the fluorescence is suppressed
due to direct energy transfer from the fluorophore to the
quencher. Consequently in a closed hairpin configuration the beacon is
not fluorescent, whereas in the open configuration it becomes
fluorescent. This property leads to many interesting applications for
molecular beacons in biology or physics.

\smallskip
Biological applications use the possible assembly of a portion of the
single strand which forms the loop with another DNA strand which is
complementary to the loop's sequence. The assembly of the single
strand of the loop with another strand to make a double helix is only
possible when the hairpin is open because double-stranded DNA is very
rigid. Therefore, when the assembly occurs, the fluorescent signal is
restored \cite{BONNET99b}. This technique provides very sensitive
probes of the sequences which are complementary to the loop. Using
this idea, it has been suggested that DNA beacons could be used in
vivo to detect the single stranded RNA which is synthesized during the
transcription of genes. This could allow the recognition of cancer
cells by targeting some genes which are heavily transcribed in those
cells \cite{PENG,SANTANGELO}.

\smallskip
Physical applications exploit the high reproducibility of the
hairpins' self-assembly process which makes it possible to build
molecular memories read by detecting the fluorescence \cite{TAKINOUE}
or devices capable of performing molecular computation
\cite{SAKAMOTO}.

\smallskip
Understanding the DNA hairpin self-assembly process at the mesoscopic
scale is possible because molecular beacons allow accurate monitoring
of the opening and closing steps.  The ``melting profile'' of the
stem, induced by heating, can be recorded accurately versus
temperature and the autocorrelation function of the fluorescence can
be used to extract the kinetics of the opening/closing
fluctuations. Measurements have been made \cite{BONNET98} for
different loop lengths and different bases in the loop. They provide a
complete set of data which can be compared to the results of a
theoretical analysis in order to determine the basic mechanisms
controlling the properties of DNA hairpins.  This is the primary aim
of the study presented here. It should be noted however that our
results, because of their strong sensitivity to the properties of the
loop, turn out to have implications which extend beyond the properties
of hairpins as such. The detailed comparison of experimental data with
the results of various loop models enhances our ability to model
single strands of DNA and RNA.

\section{The model}
\label{sec:model}

The closing of a DNA hairpin has some similarity with the folding of a
protein in the sense that it is an evolution from a random chain to a
geometrical shape which is stabilized by weak bonds established
between some of its components, here the bases of the stem. The
full process is quite complex because it involves the precise
positioning of a large number of atoms in space 
to form the strands of the stem.  However one may reasonably argue
that, in order to understand experimental observations such as the
fluctuations of a beacon, one does not need to know all the details of
the process. A simple view is to consider the DNA strand as a polymer
chain. Then it should be possible to combine known models for the stem
with a polymer model for the loop. This has been done in an approach
that uses the simplest possible model for the stem \cite{KUZNETSOV},
an Ising model in which the bases are either closed or open, and a
semiflexible polymer model for the loop. This approach gave
interesting results, in particular concerning the estimation of the
persistence length of single-stranded DNA. However it has the drawback
that the description of the stem is very rough and relies on empirical
parameters, such as the entropy change involved in the closing of two
bases, which cannot be justified within the model and have to be
fitted. Moreover, as the Ising model of the stem ignores all
geometrical parameters, such as the distance between the bases linked
to the loop, the matching between the models of the stem and the loop
has to be crude.  A further aspect which is not satisfactory in such
an approach is that it uses two different models for the stem and the
loop while both belong to the same DNA single strand, and should be
described in the same framework. This is what we are doing in the
present study.  It should of course be clear that the pairing of bases
in the stem leads to additional phenomena which do not occur in the
loop and must be taken into account in order to complete the model.
Last, but by no means least, we would like to argue that a model with
continuous degrees of freedom is more apt to describe the end-to-end
distance, which is a natural ``reaction coordinate'' measured by the
fluorescence signal.

\begin{figure}[h!]
  \includegraphics[width=8cm]{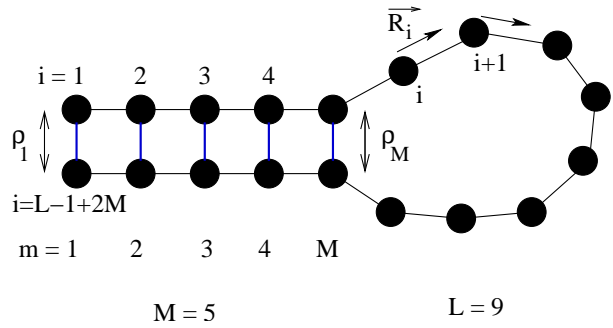}
\vskip .3truecm
\caption{ 
  A schematic picture of the model to define some notations. The
  hairpin is made of a stem of $M$ base pairs and a loop of $L$
  segments, i.e. $L-1$ bases. The bases along the strand are labeled
  by an index $i$ ranging from $1$ to $L-1+2M$. The variables $\rho_m$
  represent the distances between the bases forming base pairs $m$ in
  the stem. The {\em stretching} of the base pair distance is denoted
  by $y_m$ so that $\rho_m = y_m + d$, where $d$ is the equilibrium
  distance between the bases in a DNA double helix.
}
\label{fig:model}
\end{figure}

A schematic picture of our hairpin model is shown in
Fig.~\ref{fig:model}. It consists of a stem of $M$ base pairs and a
loop with $L$ segments, i.e.\ $L-1$ bases so that the single strand
which forms the hairpin has a total of $2M + L-1 $ bases or $N = 2M +
L-2$ segments.  This single strand can be described by different
polymer models. The dependence of our results on the particulars of
the polymer model will be discussed in Sec.~\ref{sec:discussion}.
For the moment, let us consider only one of them as the basic model of
our study, the Kratky-Porod (KP) \cite{KratkyPorod} model, also known
as the wormlike chain (WLC) in its continuum version
\cite{WilhelmFrey}; for DNA hairpins which have short loops and a very
short persistence length the original discrete version is more
appropriate. The Kratky-Porod model considers the chain of bases as
made of rigid segments of length $\ell$. The orientation of a segment
in three-dimensional space is defined by a vector $\vec{R}_i$ of
unit length, lying along segment $i$, as shown in
Fig.~\ref{fig:model}. Therefore the end-to-end distance of this chain,
labeled $\rho_1$ in Fig.~\ref{fig:model} is
\begin{equation}
   \rho_1 = \left| \ell \sum_{i=1}^{N} \vec{R}_i \right| \; .
\end{equation}
The energy of the KP model is
\begin{equation}
\label{eq:e1}
  E_1(N) = - \epsilon \ell^2 \sum_{i=1}^{N-1} \left( \vec{R}_i
\cdot \vec{R}_{i+1} - 1 \right) \; ,
\end{equation}
where $\epsilon \ell^2$ measures the energy that is necessary to bend
the polymer at a joint, i.e.\ it is a parameter that measures the
rigidity of the strand.

To complete the description of the hairpin we must also add the
interactions which may take place within the stem when
base pairing occurs. We use an approach based on the PBD model for DNA
melting \cite{PB,DPB} by adding to the polymer model the energy
contribution
\begin{align}
\label{eq:hstem}
  E_2(M) =&  
D \sum_{m=1}^{M} \Big\{ \Big[ \exp(- \alpha y_m) - 1 \Big]^2 - 1
\Big\} \nonumber \\
& +
\dfrac{1}{2} K \sum_{m=2}^{M} 
 \exp[-\zeta (y_m + y_{m-1})] \;
\big(y_m - y_{m-1}\big)^2 \nonumber \\
&\equiv \sum_{m=1}^{M} V(y_m) + \sum_{m=2}^{M}  W(y_m,y_{m-1}) \quad,
\end{align}
where $D, \alpha, K, \zeta$ are constant parameters and $y_m = \rho_m
- d$ denotes the deviation of the distance $\rho_m$ between two bases
in pair $m$ from its equilibrium value, $d$, in the double helix.  In
other words $y_m$ is the stretching of the $m^{\mathrm{th}}$ base pair
in the stem, and is a function of the vectors $\vec{R}_i$ which define
the geometrical shape of the strand. The potential energy of the stem
includes Morse potentials $V(y)$ describing the pairing energy between
two complementary bases.  The Morse potentials describe an effective
interaction which includes the attractive contribution of the hydrogen
bonds between the complementary bases and the repulsion coming from
the charged phosphate groups on the strands.  The other important
energy terms in the stem are the stacking interactions between
consecutive bases, described by the nonlinear potential $W(y,y')$.  In
fact stacking energies are also present in an implicit form in the
polymer model of the strands since the flexibility of a singe strand
of DNA is affected by the interactions between the bases which are
part of the nucleotides. In the stem however the stacking energy
increases because of the geometrical constraints of double helix
packing. The base pair plateaux are piled on top of each other and
interact strongly due to the overlap of their $\pi$ electrons. If one
of the two adjacent pairs is open the double helix packing disappears
and the prefactor $\exp[-\zeta(y_m + y_{m-1})]$ vanishes
\cite{differencePB}.  It is the geometrical constraint which allows us
to use the scalar variable $y_m$ to describe the base pair status in
the stem. In this geometry the displacement of the bases is
essentially orthogonal to the stem axis and therefore the stretching
of the individual base pairs provides a mesoscopically acceptable
description of the stem's state.

Our choice of $E_2(M)$ is based on the PBD model which has been 
widely tested for DNA melting \cite{PB,DPB} but other expressions are
certainly possible, provided they properly describe the physics of the
molecule. The potential between the bases has to include a stong
repulsion when the bases approach each other ($y_m < 0$) and the force
has to tend to zero (constant potential) at large $y_m$. The Morse
potential has the proper qualitative shape. Similarly expressing
$W(y_m,y_{m-1})$ by a harmonic interaction with an effective coupling
constant $K \exp[-\zeta (y_m + y_{m-1})] $ is a simple way to describe
the decay of the stacking which is expected when base pairs
open. Studies of variants of the PDB model show that different
expressions preserving the same qualitative properties imposed by the
physical constraints lead to quantitative changes in the results 
which are also obtainable - within the accuracy of experimental 
observations - by varying the PDB model parameters.

It should be noted that expression (\ref{eq:hstem}) imposes a
priori the bases which can be linked by a pairing potential. In other
words it assumes that the base sequences in the terminal regions of
the strand are such as to guarantee full pairing in the closed
hairpin. Mismatches are thus not allowed in the model. We do not
expect them to play a significant role in the physical system, at
least in the case of the short stems under consideration, since the
relative energetic cost of a mismatched configuration would be high.

\smallskip
The potential energy of the hairpin is $E = E_1(N) + E_2(M)$.

\section{Thermodynamic properties}
\label{seq:thermodynamics}

\subsection{Constrained partition function and free energy: principle of
  the derivation.}

The fluorescence of DNA beacons is determined by the distance $\rho_1$
between the two ends of a strand, which carry the fluorophore and 
the quencher. In order to analyze the experiments we must therefore
determine the probability ${\cal P}_N(\rho_1)d\rho_1$ that a strand
will have an end-to-end distance in the interval
$(\rho_1,\rho_1+d\rho_1)$; this is - within a normalization
factor - identical to the constrained configuration partition function
\begin{equation}
\label{eq:configpart}
{\cal Z}_N(\rho_1) = \int \prod_N d \Gamma_N \; \; \delta\left(
\left| \ell \sum_{i=1}^{N} \vec{R}_i \right| - \rho_1
\right) e^{ - \beta E(\Gamma_N) }
\end{equation}
obtained by integrating the Boltzmann weight 
over the configuration variables
symbolically denoted by $\Gamma_N$ for a DNA strand of $N$ monomers
under the constraint of fixed end-to-end distance imposed by the Dirac
delta function. The normalized probability density function
\begin{equation}
\label{eq:proba1}
{\cal P}_N(\rho_1) = \frac{{\cal Z}_N(\rho_1)}{{\cal Z}_N^0} \; ,	
\end{equation}
is obtained by dividing (\ref{eq:configpart}) by the unconstrained 
partition function
\begin{equation}
\label{eq:uncpart}
	{\cal Z}_N^0= \int \prod_N d \Gamma_N \; \; 
	e^{ - \beta E(\Gamma_N) }\quad.
\end{equation}

\medskip
The calculation of ${\cal Z}(\rho_1)$ for a hairpin is complicated by the
presence of interactions within the stem because they involve the {\em
relative positions} of two segments of the polymer. In order to
proceed let us start from the $L$ segments forming the loop. Their
study is simpler because they form an ordinary polymer, which, in our
case, is described by the KP model. The configuration partition
function of the loop is $Z_L^0$ and the probability that the two ends
of the loop are at distance $\rho_M$, the distance between the two
bases at the end of the stem connected to the loop, is
\begin{equation}
\label{eq:probaL}
  P_L(\rho_M) = Z_L(\rho_M) / Z_L^0 \; .
\end{equation}
In contrast
to the expressions of Eq.~(\ref{eq:proba1}) which refer to the full 
hairpin, the corresponding
terms in Eq.~(\ref{eq:probaL}) refer to
an ordinary polymer without the additional constraints
imposed by
the stem. Their derivation is discussed in the
subsection~\ref{subsec:KP}.
To stress this distinction we have used a script notation
${\cal Z}$, ${\cal P}$ for quantities that
cannot be obtained from standard polymer theory.

\smallskip
Now that the loop is characterized, let us derive the partition
function of the hairpin by successively adding the segments which form
the stem, one segment at a time. If we start from the loop and add one
monomer at each end, we have built one segment of the stem. The
distance between the new ends of the strand is now $\rho_{M-1}$. Using
the notation of Eqs.~(\ref{eq:proba1}) or (\ref{eq:probaL}), for this
extended polymer consisting of $L+2$ monomers, we have
\begin{equation}
  {\cal Z}_{L+2}(\rho_{M-1}) = {\cal P}_{L+2}(\rho_{M-1}) \; 
{\cal Z}_{L+2}^0 \; .
\end{equation}
In order to evaluate ${\cal P}_{L+2}(\rho_{M-1})$, let us introduce a {\em
conditional probability} $S(\rho'|\rho)$ that, if a polymer of $p$
monomers has its ends at distance $\rho$, a polymer of $p+2$ monomers,
obtained by adding one monomer at each end of the previous one, has
the distance $\rho'$ between its ends. For the polymer alone i.e.\
without the contribution of the energy $E_2$ in the stem, this
conditional probability is such that
\begin{equation}
\label{eq:PSP}
  P_{p+2}(\rho') = \int_{0}^{\infty} S(\rho'|\rho) \;
   P_p(\rho) \; d \rho \; .
\end{equation}
In the presence of the energy terms $E_2$ within the stem, the
conditional probability $S(\rho'|\rho)$, determined by the properties
of the polymer alone, must be corrected by a Boltzmann factor
containing the potential energy terms $V(\rho) + W(\rho',\rho)$ due to
base pairing and stacking interactions.  Accordingly, the hairpin with
a single pair of complementary bases will satisfy
\begin{align}
\label{eq:Plplus2}
  {\cal P}_{L+2}(\rho_{M-1}) = \int_{0}^{\infty}
d \rho_M \; &e^{- \beta [V(\rho_M) + W(\rho_{M-1},\rho_M)]}
 \nonumber \\
 & \times S(\rho_{M-1}|\rho_M) \; P_L(\rho_M) \; .
\end{align}
The process can be iterated to add the remaining segments of the
stem. The advantage of this progressive buildup of the stem is that it
explicitly introduces the distances between the bases that pair in the
stem in the calculation, allowing us to include the proper statistical
weights arising from pairing and stacking energies in the stem.

Once all the stem segments and stem energy terms have been included we 
obtain
\begin{align}
\label{eq:Zrho1}
  {\cal Z}_N(\rho_1) = {\cal Z}_N^0 \int & d\rho_2  
  \ldots  \int d\rho_M \;
e^{- \beta V(\rho_1)} \nonumber \\ 
& \times e^{- \beta [V(\rho_2) + W(\rho_1, \rho_2)]} \times \ldots \nonumber\\
& \times e^{- \beta [V(\rho_M) + W(\rho_{M-1}, \rho_M)]} \; \nonumber \\
& \times S(\rho_1|\rho_2) \;  
\ldots S(\rho_{M-1}|\rho_M)
  \nonumber \\
& \times P_L(\rho_M) \; .
\end{align}
This expression gives the constrained partition function of the
hairpin in terms of properties of the polymer forming the strand,
$P_L(\rho)$ and $S(\rho'|\rho)$. It is therefore valid for any
polymer model, provided one can derive these two probability
distributions for the model of interest. The constrained partition
function ${\cal Z}_N(\rho_1)$ defines an effective 
free energy ${\cal F}(\rho_1) = -
k_B T \ln {\cal Z}_N(\rho_1)$ for the hairpin having the distance $\rho_1$
between its ends, i.e.\ it gives the free energy landscape using
$\rho_1$ as the relevant coordinate.  The result appears as a
$(M-1)$-dimensional integral but - like any transfer integral - it can
actually be computed by a sequence of $M-1$ one-dimensional
integrations.
Performing first the integration over $\rho_M$, we get a function of
$\rho_{M-1}$; next, the integration over $\rho_{M-1}$ gives a function
of $\rho_{M-2}$, and so on, until the last integration over $\rho_2$
which gives the desired constrained partition function. Therefore, the
calculation of $Z_N(\rho_1)$ is a relatively straightforward numerical
task since $P_L(\rho)$ and $S(\rho'|\rho)$ can be derived from
appropriate polymer models.

\subsection{The properties of the Kratky-Porod model, and the 
effective Gaussian approximation}
\label{subsec:KP}

In order to proceed further with the calculation of ${\cal F}(\rho_1)$ we need
expressions of $P_L(\rho)$ and $S(\rho'|\rho)$ for the polymer
model chosen to describe the DNA strand, i.e.\ the Kratky Porod (KP)
model having an energy given by Eq.~(\ref{eq:e1}). This model has been
widely studied in the continuum limit, known as wormlike chain (WLC)
where the energy tends to
\begin{equation}
  E'_1 = \dfrac{\kappa}{2} \int_0^{\Lambda} dx \left| \dfrac{\partial
  \vec{R}}{\partial x} \right|^2
\end{equation}
for a given chain length $\Lambda = L \ell $, in the limit $L \to
\infty$, $\ell \to 0$, provided $\epsilon \ell^3 \to \kappa$, the
continuum chain stiffness. 

However, the probability distribution function $P(\rho)$ 
obtained in the continuum limit \cite{WilhelmFrey,HAMPRECHT,SamSinh,StepSch}
is not appropriate for DNA hairpins for which the loops may not
be longer than a few persistence lengths of single-stranded DNA and
the persistence length itself hardly exceeds the monomer distance.  In
this case the continuum limit becomes a priori questionable and the
discrete expression of Eq.~(\ref{eq:e1}) should be preserved. The
partition function $Z_L^0$ of a polymer of $L$ segments is readily
calculated as
\begin{equation}
  Z_L^0 = \int d\Omega_1 \ldots d\Omega_L e^{-\beta E_1(L)} = 4 \pi
  \left[ 4 \pi e^{-b} i_0(b) \right]^{L-1} \; ,
\end{equation}
where $b = \beta \epsilon \ell^2$ and $i_0(b) = \sinh(b) / b$ is the
modified Bessel function of zeroth order. The 
mathematical equivalence of this model with the classical Heisenberg
ferromagnetic chain \cite{Fisher} can be used to show that the
orientational correlations between different segments have the form
\begin{equation}
  \langle \vec{R}_r \cdot  \vec{R}_s \rangle =
e^{ - |r - s| \ell / \lambda} \; ,
\end{equation}
with a persistence length
\begin{equation}
  \label{eq:persist}
\lambda = -  \dfrac{\ell }{\ln[i_1(b)/i_0(b)]} = 
- \dfrac{\ell}{\ln[\coth (b) - 1/b]} \; ,
\end{equation}
where $i_1(b) = [b \cosh(b) - \sinh(b) ]/b^2$ is the modified Bessel 
function of first order.
For the discrete KP model, the end-to-end distribution function can be
computed numerically from its Fourier transform, which can be
expressed \cite{Marko} as the leading matrix element
\begin{equation}
\label{eq:eteKPq}
	P_N^{KP}(\vec q) = \left( {\bf F}^N \right)_{00}
\end{equation}
of the $N$th power of a symmetric matrix ${\bf F}$ whose 
elements are given by

\begin{eqnarray}
\nonumber
	F_{ll'}(q) &= & \frac{1}{2} \left[(2l+1)(2l'+1)
	{\hat i}_l(b) {\hat i}_{l'}(b)  \right]^{1/2} \\
	    &  & \sum_{k=|l-l'|,k+l+l'=2r}^{l+l'}  (2k+1) (-i)^k 
	    \frac{1}{r+1/2} \nonumber \\
	& &    \frac{\Psi(r-k)\Psi(r-l)\Psi(r-l')}{\Psi(r)} 
	    j_k(q)  \quad,
\label{eq:eteKPmatelem}	    
\end{eqnarray}

where 

\[
	\Psi(n) = \frac{\Gamma(n+\frac{1}{2} ) }  {\Gamma(n+1)\Gamma(\frac{1}{2})} =
	\prod_{j=1}^{n}\left( 1 - \frac {1}{2j}\right) \quad,
	\]
\noindent	
$j_k(q)  $ is the spherical Bessel function	of $k$th order, and 
$ {\hat i}_l(b) = i_l(b)/i_0(b)$. 
In practice one can obtain numerically accurate results
even for short stiff polymers ($L=10$, $\lambda = 0.8 L \ell$ for instance)
by  summing no more than 8 terms in  (\ref{eq:eteKPmatelem}). 
Furthermore, since only the leading matrix element is 
required, direct matrix multiplication is quite efficient.

\smallskip
The derivation of the conditional probability $S(\rho'|\rho)$ is even
more demanding than the calculation of $P(\rho)$ and we have not
been able to obtain it for the KP model. Fortunately however, in the
case of weak chain rigidity, there is a way to go around this
difficulty because,  as shown in 
appendix \ref{app:srhorhop}, 
the conditional probability $S(\rho'|\rho)$ can be
calculated exactly for a Gaussian chain, made of orientationally
uncorrelated links such that the probability for any segment to lie
along a vector $\vec{\Delta}$ is proportional to $\exp(-
|\vec{\Delta}|^2 / 4 \tau^2 )$. It is given by
\begin{equation}
\label{eq:Sgauss1}
 S(\rho'|\rho) = \sqrt{\dfrac{1}{2 \pi \tau^2}} \; \dfrac{\rho'}{\rho}
 e^{-(\rho'^2 + \rho^2)/8 
 \tau^2}
\sinh\left( \dfrac{\rho' \rho}{4 \tau^2} \right) \; .
\end{equation}

\begin{figure}
\begin{center}
$L = 24 $ \\
 \includegraphics[width=8cm]{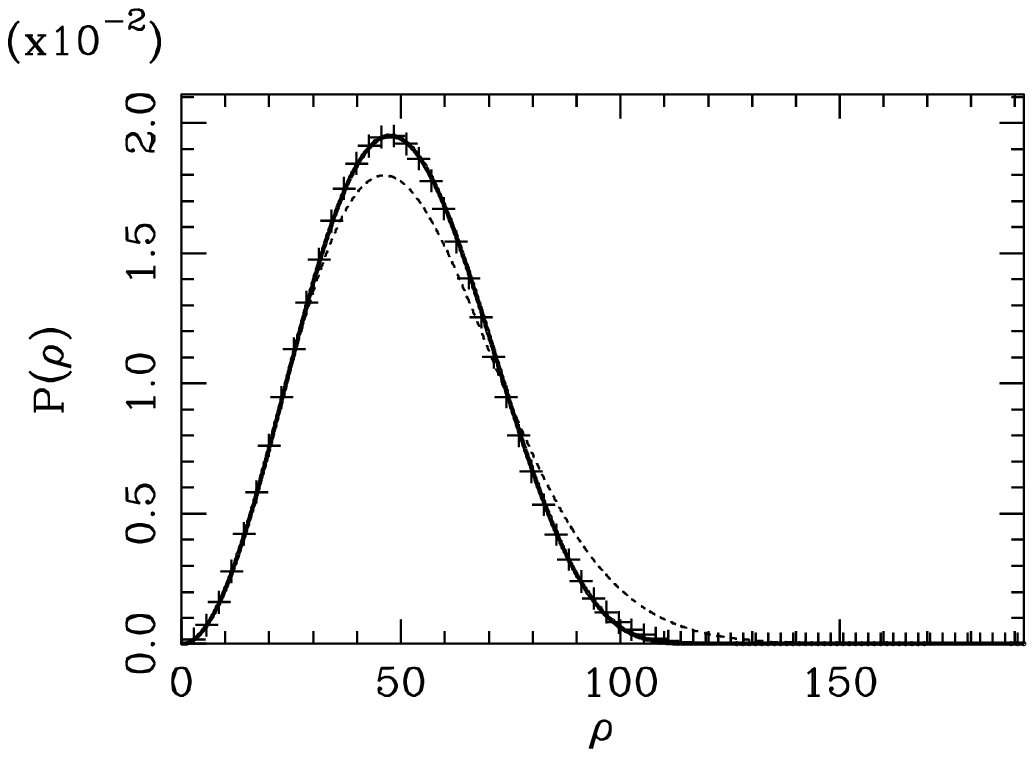} \\
$L = 14$ \\
 \includegraphics[width=8cm]{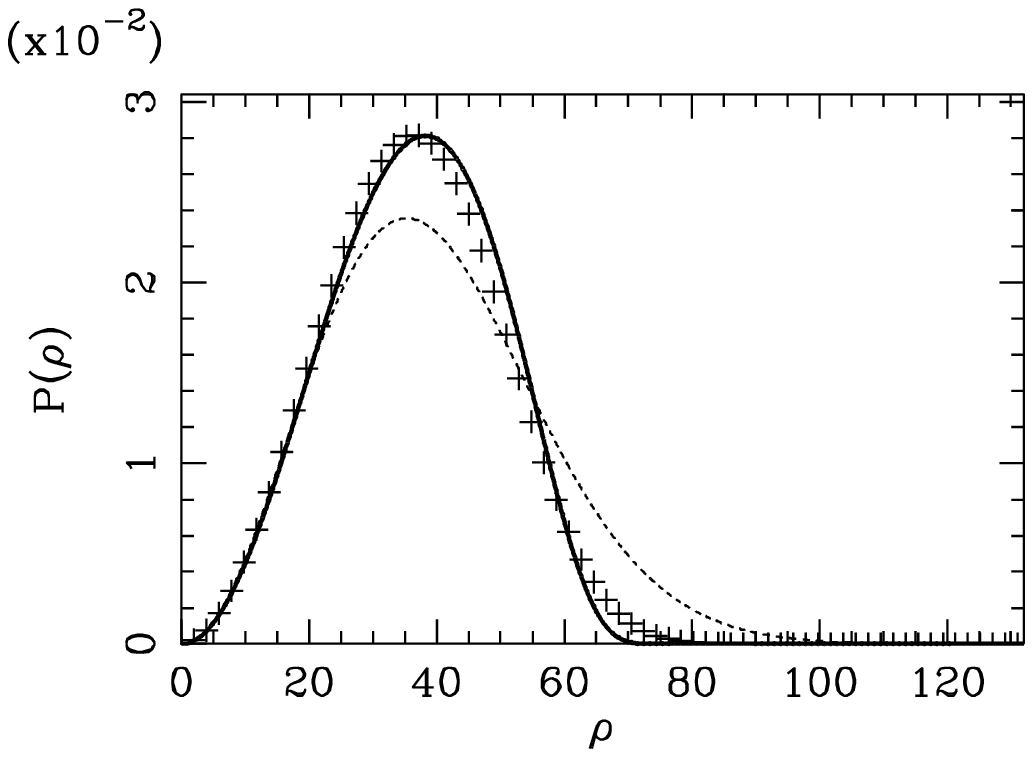}
\end{center}
\caption{Comparison between the Kratky Porod
  distribution function $P^{KP}_L(\rho)$ (full line) and the Gaussian
  approximation $P^G_L(\rho)$ (dashed line) for two values of $L$. 
The crosses
  show the distribution function $P_L(\rho)$ obtained
  by starting from the Kratky Porod distribution $P^{KP}_{L-2}(\rho)$
  and computing the probability distribution of the end-to-end
  distance of a polymer extended by two units, using the conditional
  probability $S(\rho'|\rho)$ according to Eq.~\ref{eq:PSP}.
The parameters for the KP model are $\epsilon = 0.0016\;$eV/\AA$^2$,
  $\ell = 6\;$\AA, and the calculation has been made for $T=300\;$K,
  giving $b = \beta \epsilon \ell^2 = 2.26$ and $\lambda = 10.8\;$\AA.
}
\label{fig:probafit}
\end{figure}

\bigskip
The Gaussian probability function $P^G(\rho)$ can be used to
approximate the end-to-end distribution function of the KP chain 
$P^{KP}(\rho)$ by choosing a temperature dependent value of its
parameter $\sigma^2 = L \tau^2$ so that the average square of the
end-to-end distance of a Gaussian chain with $L$ segments $\langle
\rho^2 \rangle = L\; \ell^2 = 6 \sigma^2 = 6 L \tau^2$ matches the
average value of $\langle \rho^2 \rangle$ for the KP
chain
\begin{equation}
  \langle \rho^2 \rangle =  L\; \chi
\end{equation}
with
\begin{equation}
  \chi = \ell^2 \dfrac{1 + \coth (b) - 1/b}{1 - \coth (b) + 1/b} \; .
\end{equation}
To get a Gaussian approximation for the KP chain we must therefore select
\begin{equation}
\label{eq:tau2}
  \tau^2 = \dfrac{\chi}{6} \; ,
\end{equation}
Figure \ref{fig:probafit} shows that the Gaussian approximation is
fairly good for $L=24$ and becomes poor for $L=14$. However we do not
actually need to use $P^G(\rho)$. The advantage of the Gaussian
approximation is that it provides the basis for an approximate
expression of $S(\rho'|\rho)$ given by Eqs.~(\ref{eq:Sgauss1}) and
(\ref{eq:tau2}), which can be used in order to compute $P_{L+2}$ from
$P_L$, according to Eq.~(\ref{eq:Plplus2}), by providing for 
$P_L$ the numerical result $P^{KP}_L(\rho)$, i.e.\ a value which
is very accurate.
In this approach the error introduced by the Gaussian approximation
only affects the {\em variation} of $P(\rho)$ when the polymer is
extended.  Figure \ref{fig:probafit} shows that, even for a short loop
$L=14$, for which the Gaussian approximation is poor, the comparison
between the approximated expression of $P_{L+2}$ and the accurate
numerical value $P^{KP}_{L+2}$ is quite good, and becomes
excellent for longer loops ($L=24$). This gives us all the ingredients
that we need to compute the constrained partition function of the
hairpin ${\cal Z}_N(\rho_1)$ according to Eq.~(\ref{eq:Zrho1}).

\subsection{First results.}
\label{subsec:results}

Before discussing all the results in Sec.~\ref{sec:discussion} it is
useful to consider an example which illustrates the thermal properties
of hairpins and introduces some quantities which will turn out to be
relevant in the next section on kinetic properties.

\begin{figure}[h!]
\begin{center}
 \includegraphics[width=8cm]{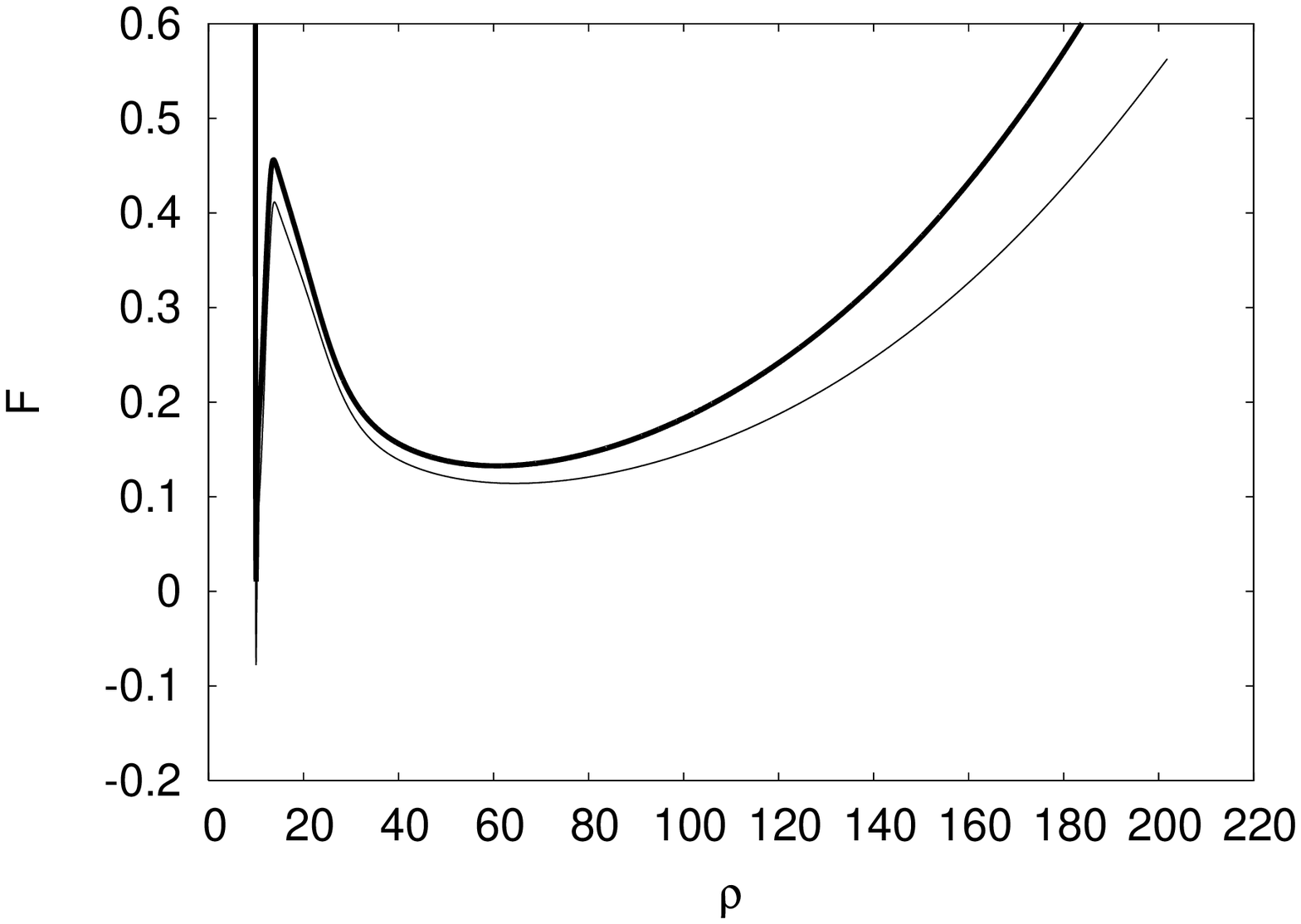} \\
 \includegraphics[width=8cm]{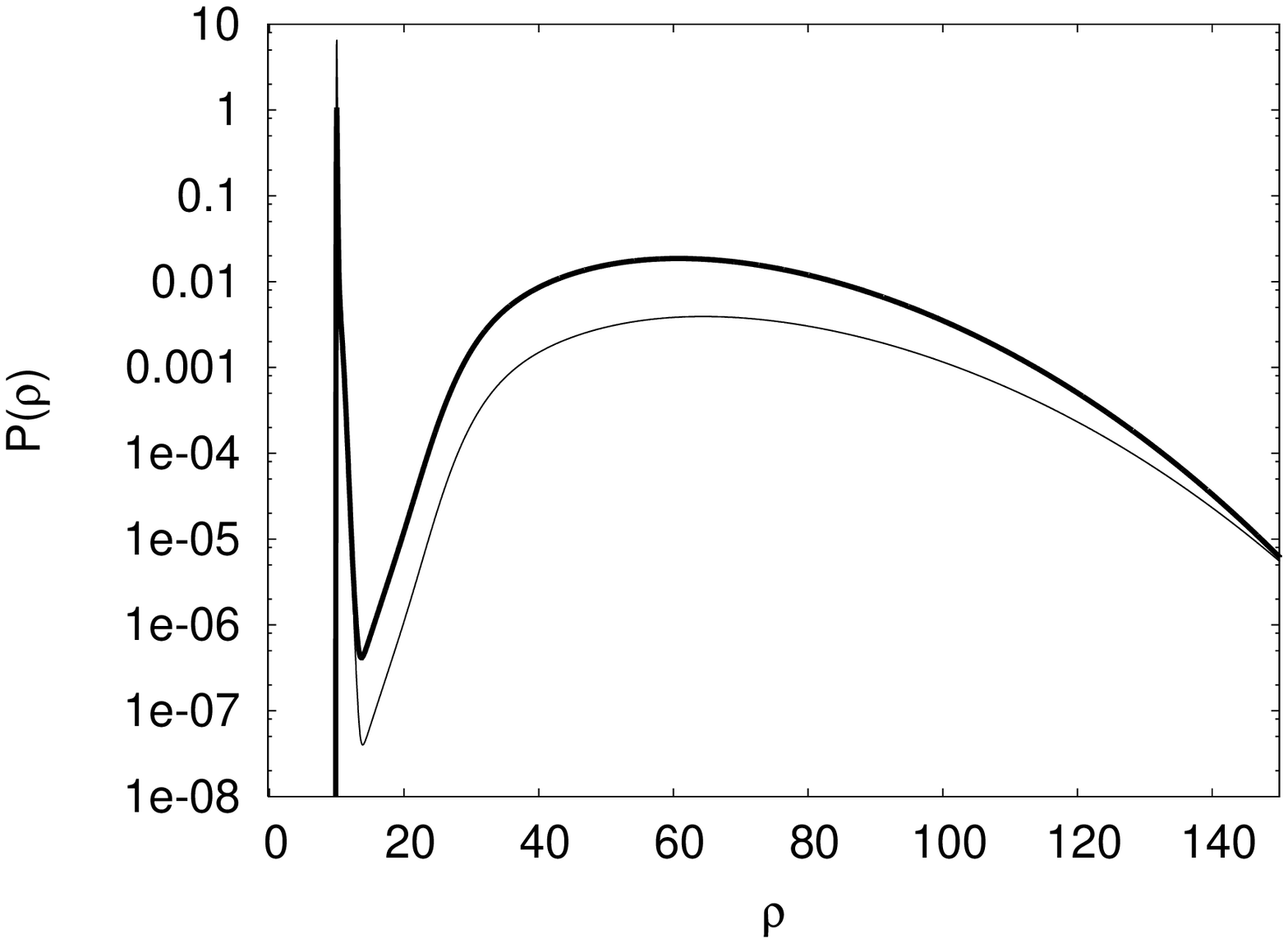}
\end{center}
\caption{Behavior of the hairpin model at two different temperatures,
  $T=300\;$K (thin line) and $T = 350\;$K (thick line), 
for $L=24$.
Top figure: free energy 
versus $\rho_1$, bottom figure: probability density
  $P(\rho_1)$ in logarithmic scale. The parameters of the model are
$D = 0.16\;$eV, $\alpha = 6.9\;$\AA$^{-1}$, $K=0.125\;$eV/\AA$^2$,
  $\zeta=0.10\;$\AA$^{-1}$, $\epsilon=0.0016\;$eV/\AA$^2$, $\ell = 6\;$\AA.
}
\label{fig:fenerg}
\end{figure}

\begin{figure}[h!]
  \includegraphics[width=8cm]{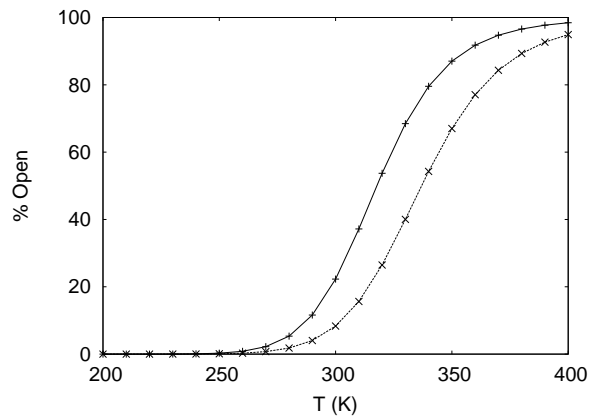}
\caption{Variation versus temperature of the percentage of open hairpins
  for two different loop lengths $L=24$ (full line) and $L=14$ (dotted
  line).
The parameters of the model are $M=5$, $D=0.16\;$eV, 
$\alpha = 6.9\;$\AA$^{-1}$ $K=0.125$, 
$\zeta= 0.10\;$\AA$^{-1}$, 
$\epsilon = 0.0016\;$eV/\AA$^2$, $\ell = 6\;$\AA.
}
\label{fig:opnlength}
\end{figure}

Figure \ref{fig:fenerg} shows the effective free energy ${\cal F}(\rho_1)$
for a hairpin with $M=5$ base pairs in the stem and $L=24$ segments in
the loop, at $300\;$K and $350\;$K, and the corresponding probability
distributions ${\cal P}_N(\rho_1)$. As shown below, these two temperatures
are on both sides of the opening temperature $T_m$ of this
hairpin. However Fig.~\ref{fig:fenerg} shows that ${\cal F}(\rho_1)$ and
${\cal P}_N(\rho_1)$ maintain the same qualitative shape at both
temperatures. ${\cal F}(\rho_1)$ has a narrow well around $\rho_1 =
10\;$\AA\, which is the equilibrium distance between the bases in a
DNA double helix; the narrow well is separated from a broad secondary
minimum at larger $\rho_1$ by a fairly sharp maximum. The probability
density ${\cal P}_N(\rho_1) = {\cal Z}_N(\rho_1)/{\cal Z}_N^0 = \exp[-\beta
{\cal F}(\rho_1)]/{\cal Z}_N^0$ exhibits two peaks. The peak around $\rho_1 =
10\;$\AA\ corresponds to the closed state of the hairpin, while the
broad maximum at large $\rho_1$ corresponds to the open
configurations. This shape of ${\cal P}_N(\rho_1)$ points out that, at any
temperature, the open and closed forms of the hairpin coexist. The
opening ``transition'' is only a shift of the equilibrium from one
temperature regime where the closed configurations dominate to another
where the open states are the majority. This is not surprising since
even an approximate phase transition should not be expected in a small
system such as a DNA hairpin. Therefore, in order to provide a measure
of the opening of the hairpin we have to compute the {\em fraction of
open states}, which can be obtained from the probability distribution
${\cal P}_N(\rho_1)$ by defining as closed the states for which
$\rho_1 \le \rho^{\star}$ and open those for which $\rho^{\star} <
\rho_1 < \rho_{\mathrm {max}}$, where $\rho^{\star}$ is the value of
$\rho_1$ corresponding to the minimum of ${\cal P}_N(\rho_1)$ (i.e.\ the
maximum of ${\cal F}(\rho_1)$) and $\rho_{\mathrm {max}} = N \ell$ is the
maximum distance between the ends of the DNA strand, determined by the
length of the strand. The respective probabilities to find the hairpin
in the closed and open configurations are thus
\begin{equation}
  \label{def:pcpo}
p_c = \int_0^{\rho^{\star}} {\cal P}_N(\rho_1) d\rho_1
\qquad
p_o = \int_{\rho^{\star}}^{\rho_{\mathrm {max}}} {\cal P}_N(\rho_1)
  d\rho_1 \; .
\end{equation}
Since ${\cal P}_N(\rho_1)$ is normalized, i.e.\ $p_c + p_o = 1$, $p_o$
also represents the fraction of open configurations at a given
temperature.  Performing such a calculation as a function of
temperature gives the so-called ``melting curve'' of the DNA
hairpin. Figure \ref{fig:opnlength} shows two examples of such curves
for $L=24$ (the case illustrated in Fig.~\ref{fig:fenerg}) and a case
with a shorter loop ($L=14$). If we define $T_m$ as the temperature at
which $p_o = p_c$, we get $T_m(L=24) = 317.7\;$K and $T_m(L=14) =
337.0\;$K for the model parameters that we used in these
calculations. In the context of this paper we will also refer to $T_m$
as the opening temperature of the hairpin.

\section{Kinetics}
\label{sec:kinetics}

The derivation of the free energy ${\cal F}(\rho_1)$ allows us to go beyond
the analysis of the equilibrium properties of the hairpins because it
exhibits the characteristic shape of a system evolving between 3
states $C \leftrightarrows T^\star \leftrightarrows O$, the closed $C$
and open $O$ states associated{, respectively, with the minima of
${\cal F}(\rho_1)$ and an unstable transition state $T^\star$ corresponding
to the intermediate maximum. This suggests that the multidimensional
dynamics of the opening and closing of the hairpins can be viewed as a
reduced problem of reaction kinetics. If the system is strongly
coupled to its environment, the dynamics of the molecule has no memory
of its velocity so that it is well described by a diffusion on the
free energy surface ${\cal F}(\rho_1)$. For the large molecular units
involved in the opening/closing of DNA hairpins this is a reasonable
assumption. Studying the kinetics of hairpin fluctuations is thus
reduced to the calculation of a first passage time in a diffusion
controlled process \cite{Schulten,Szabo}.

\begin{figure}[h!]
\begin{center}
  \includegraphics[width=4cm]{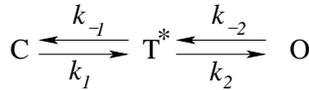}
\end{center}
\caption{Schematic of the reaction kinetics view of the hairpin
  opening/closing, and definition of the reaction rate constants of
  the processes involved.}
\label{fig:kinetics}
\end{figure}

In the language of chemical reaction kinetics, if we denote the
 concentrations in the three states by $p_c$, $p^{\star}$, $p_o$ ,
 respectively, and use the kinetic constants defined in
 Fig.~\ref{fig:kinetics}, the temporal evolution of the reactants is
 described by
\begin{align}
  \label{eq:kin1}
\dfrac{d p_c}{dt} &= - k_1 p_c + k_{-1} p^{\star} \\
\dfrac{d p_o}{dt} &= k_2 p^{\star} - k_{-2} p_o \\
\dfrac{d p^{\star}}{dt} &= k_1 p_c + k_{-2} p_o - (k_{-1} + k_{2})
p^{\star}   \quad.
\end{align}
Under the standard assumption of rapid intermediate state dynamics,
there is no variation of the concentration $p^{\star}$ on the time
scale of the diffusive motion which controls barrier crossing.  The
condition $d p^{\star}/dt = 0$ implies
\begin{equation}
p^{\star} = \dfrac{k_1 p_c + k_{-2} p_o}{k_{-1} + k_2}
  \label{eq:pstar}
\end{equation}
and
allows us to eliminate the concentration of the transition state
from the equations, leading to
\begin{align}
  \label{eq:kin2}
\dfrac{d p_c}{dt} &= - \dfrac{k_1 k_2}{k_{-1}+k_2} p_c + 
\dfrac{k_{-1} k_{-2}}{k_{-1} + k_2} p_o \equiv - k_f p_c + k_r p_o\\
\dfrac{d p_o}{dt} &= - \dfrac{k_{-1} k_{-2}}{k_{-1} + k_2} p_o +
\dfrac{k_1 k_2}{k_{-1}+k_2} p_c \equiv - k_r p_o + k_f p_c  \quad.
\end{align}

The equilibrium concentrations $\bar{p}_c$ and
$\bar{p}_o$ satisfy the relationship
$\bar{p}_c/\bar{p}_o = k_r/k_f$, 
which allows us to rewrite the inverse of the forward and 
reverse kinetic constants as
\begin{align}
\label{eq:kfkr}
k_f^{-1} &= k_1^{-1} + \dfrac{\bar{p}_c}{\bar{p}_o} k_{-2}^{-1}
\\ 
k_r^{-1} &= k_{-2}^{-1} + \dfrac{\bar{p}_o}{\bar{p}_c}
k_1^{-1} \; .
\end{align}
The ratio of 
equilibrium concentrations is given by the ratio of 
partition functions of the corresponding states
\begin{equation}
  \dfrac{\bar{p}_c}{\bar{p}_o} =
\dfrac{{\cal Z}_c}{{\cal Z}_o} \; ,
\end{equation}
where ${\cal Z}_c$ and ${\cal Z}_o$ designate the partition function
of the hairpin restricted to $\rho_1 < \rho^{\star}$ or $\rho_1 >
\rho^{\star}$  
respectively. The kinetic parameters are now expressed only in terms
of equilibrium properties and the lifetimes $k_1^{-1}$ and
$k_{-2}^{-1}$ 
of the closed and open states, which we must evaluate. 

Each of the two states corresponds to a basin of the free energy
${\cal F}(\rho_1)$, and the lifetime of the closed and open states is
therefore the first passage time of the coordinate $\rho_1$ above the
barrier that defines the boundary between the two basins for $\rho_1 =
\rho^{\star}$. The diffusion on the free energy surface is
described by the Smoluchowski equation
\begin{align}
\label{eq:FP1}
  \dfrac{\partial {\cal P}(\rho_1,t)}{\partial t} &=
- \dfrac{\partial j(\rho_1,t)}{\partial \rho_1} \\
j(\rho_1,t) &= - D_0 \left[\dfrac{\partial {\cal P}(\rho_1,t)}{\partial \rho_1}
+ \beta \dfrac{\partial {\cal F}(\rho_1)}{\partial \rho_1} {\cal P}(\rho_1,t)
\right] \; ,
\label{eq:FP2}
\end{align}
where $j(\rho_1,t)$ is the current of the probability ${\cal P}(\rho_1,t)$
that the distance between the ends of the hairpin is $\rho_1$ at time
$t$.

The diffusion coefficient $D_0$ is determined by the actual diffusive
mechanism of the elements of the DNA strand in the solvent that
surrounds the hairpin. It could in principle depend on $\rho_1$, but a
reasonable assumption in an ordinary solvent is to consider $D_0$ as a
constant. Its value sets the timescale of the opening/closing of the
hairpin.

The calculation of the first passage time $\tau$ for
Eq.~(\ref{eq:FP2}) has been made by Szabo et al.\ \cite{Szabo}; an
alternative derivation, outlined in Appendix \ref{app:calcultau} for
the sake of completeness, has been given by Deutsch
\cite{Deutsch}. The result is
\begin{equation}
\label{eq:resutau}
  \tau = \int_{\rho_0}^{\rho^{\star}} dr \dfrac{1}{D_0 p_0(r)} I^2(r) \; ,
\end{equation}
with
\begin{equation}
\label{eq:Ir}
  I(r) = \int_{\rho_0}^r d\rho \; p_0(\rho) \; ,
\end{equation}
where $\rho_0$ defines the limit of the basin of interest ($\rho_0 = 0$
for the basin corresponding to closed hairpins, $\rho_0 =
\rho_{\mathrm{max}}$ for open hairpins) and $p_0(r)$ is the
probability that $\rho_1 = r$ in the basin of interest, determined by
the free energy ${\cal F}(\rho_1)$ according to
\begin{equation}
  p_0(r) = \dfrac{e^{-\beta {\cal F}(r)}}{\int_{\rho_0}^{\rho^{\star}} d\rho
e^{-\beta {\cal F}(\rho)}} \; .
\end{equation}

From Eq.~(\ref{eq:kfkr}) we get
\begin{align}
\label{eq:kf}
  k_f^{-1} = &\int_0^{\rho^{\star}} dr \dfrac{1}{D_0 \exp[-\beta
  {\cal F}(r)]/{\cal Z}_c} I_c^2(r)  \nonumber \\ & +
  \dfrac{{\cal Z}_c}{{\cal Z}_o} \int_{\rho^{\star}}^{\rho_{\mathrm{max}}}
   dr \dfrac{1}{D_0 \exp[-\beta
  {\cal F}(r)]/{\cal Z}_o} I_o^2(r) \; ,
\end{align}
where we  denoted
by $I_c(r)$ and $I_o(r)$ the integral (\ref{eq:Ir}) computed in the
basin for closed or open states respectively. To avoid overflows in
the calculations it is convenient to rewrite those integrals by
introducing inside them the factor $\exp[\beta {\cal F}(r)]$. If we
define
\begin{align}
  \label{rq:Jr}
J(r) &= \dfrac{1}{{\cal Z}_c} \int_0^{r} d\rho e^{-\beta[{\cal F}(\rho) - {\cal
      F}(r)]} 
\qquad {\mbox{for}} \quad r < \rho^{\star} \\
J(r) &= \dfrac{1}{{\cal Z}_o}   \int_r^{\rho_{\mathrm{max}}} d\rho 
e^{-\beta[{\cal F}(\rho) - {\cal F}(r)]}
\qquad {\mbox{for}} \quad r > \rho^{\star} \; ,
\end{align}
equation (\ref{eq:kf}) gives
\begin{equation}
   k_f^{-1} = {\cal Z}_c \int_0^{\rho_{\mathrm{max}}} dr \dfrac{1}{D_0}
e^{-\beta {\cal F}(r)} J^2(r) \; ,
\end{equation}
and an equivalent expression for $k_r^{-1}$ with ${\cal Z}_o$ 
can also be obtained. 

\begin{figure}[h!]
  \includegraphics[width=8cm]{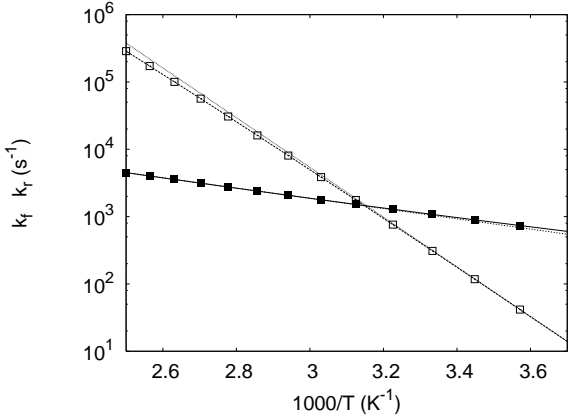}
\caption{Opening (open squares) and closing (closed squares) reaction
rates $k_f$ and $k_r$, in
  logarithmic scale, versus $1000/T$. $L = 24$. The other model
parameters are the same as for Fig.~\ref{fig:opnlength}.
The calculation has been made
  with $D_0 = 3.0\; 10^6\;$cm$^2$/s.
The full lines show fits by Arrhenius laws with $E_o = 0.73\;$eV and
$E_c = 0.15\;$eV.
}
\label{fig:topencl}
\end{figure}

Figure \ref{fig:topencl} shows the temperature dependence 
of the opening and closing times, $k_f^{-1}$ and $k_r^{-1}$ respectively. 
The values are
proportional to $D_0^{-1}$, the inverse of the diffusion coefficient
introduced in the Smoluchowski equation (\ref{eq:FP2}). Measurements
for single strands of DNA \cite{Stellwagen} give diffusion
coefficients of $1.5\; 10^6\;$cm$^2$/s. We have used the value $D_0 =
3.0\; 10^6\;$cm$^2$/s, which is a reasonable estimate for the shorter
pieces of DNA strands involved in the closing of the hairpins that we
consider. This choice leads to time scales of $k_f^{-1}$ and $k_r^{-1}$
which are in good agreement with the experiments \cite{BONNET98}.

Their temperature dependence is well fitted by Arrhenius laws
\begin{equation}
  k_f^{-1} \propto e^{\beta E_o} \qquad k_r^{-1} \propto e^{\beta E_c}
  \; .
\end{equation}
Both activation energies are positive in agreement with the
experimental observations \cite{BONNET98}. The opening activation
energy $E_o = 0.74\;$eV is very close to $M D = 0.80\;$eV which is the
energy corresponding to the breaking of the $M$ base pairs of the
stem.

\section{Discussion, role of the model of the loop.}
\label{sec:discussion}

\bigskip\bigskip
The properties of the model can be examined in the light of
experimental studies of DNA beacons which investigated the effect of
the length and composition of the loop \cite{BONNET98,GODDARD}.

\begin{figure}[!]
  \includegraphics[width=8cm]{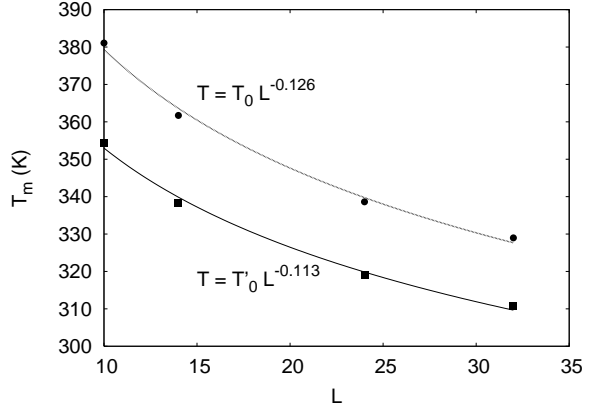}
\caption{Variation of the opening temperatures of DNA hairpins deduced
  from the model where the loop is described by a Kratky Porod model
  for $\epsilon=0.0012\;$eV/\AA$^2$ (circles) and
  $\epsilon=0.0016\;$eV/\AA$^2$ (squares). The curves show fits with
  the function $T = T_0 L^{\nu}$.
}
\label{fig:Tm}
\end{figure}

Figure \ref{fig:Tm} shows the variation of the opening temperature
$T_m$ versus the length of the loop for two values of the parameter
$\epsilon$ that governs the rigidity of the Kratky Porod model,
$\epsilon=0.0016\;$eV/\AA$^2$ giving a persistence length $\lambda =
10.82\;$\AA\ at $300\;$K ($\lambda/\ell = 1.8$) and
$\epsilon=0.0012\;$eV/\AA$^2$ giving a persistence length $\lambda =
8.05\;$\AA\ at $300\;$K ($\lambda/\ell = 1.34$). Measurements of the
persistence length for single-stranded poly(T) DNA give values in the
range $7.5$ to $13\;$\AA, depending on the salt conditions, with some
measurements leading to values as high as $40\;$\AA\ 
\cite{Smith,Rivetti}. Single-stranded poly(A) can be expected to have a
larger persistence length because adenine bases are larger than
thymines. However, for short loops it may be difficult to draw a
definite conclusion because some all-atom molecular dynamics
simulations \cite{Cuesta} show that the larger bases may be expelled
from the inside of the loop due to steric repulsions while the smaller
ones may stay inside and stack on each other. Paradoxically this could
lead to a larger flexibility for a poly(A) loop than for a poly(T). This
points out the difficulty to get reliable values of the persistence
length from experiments that do not investigate the hairpins
themselves. However the values of $\epsilon$ that we have selected are
in the expected range for single-stranded DNA, and we assume that the
larger value of $\epsilon$ corresponds to poly(A). Figure \ref{fig:Tm}
shows that, for a given loop length, $T_m$ decreases when the rigidity
of the loop increases, in agreement with experiments
\cite{GODDARD}. Moreover, as observed experimentally, the melting
temperature of the hairpins decreases with increasing loop length. For
the model we obtain $T_m \propto L^{- \nu}$ with $\nu \approx
0.12$. 

\begin{figure}[h!]
  \begin{center}
$\epsilon=0.0016\;$eV/\AA$^2$ \\
\includegraphics[width=8cm]{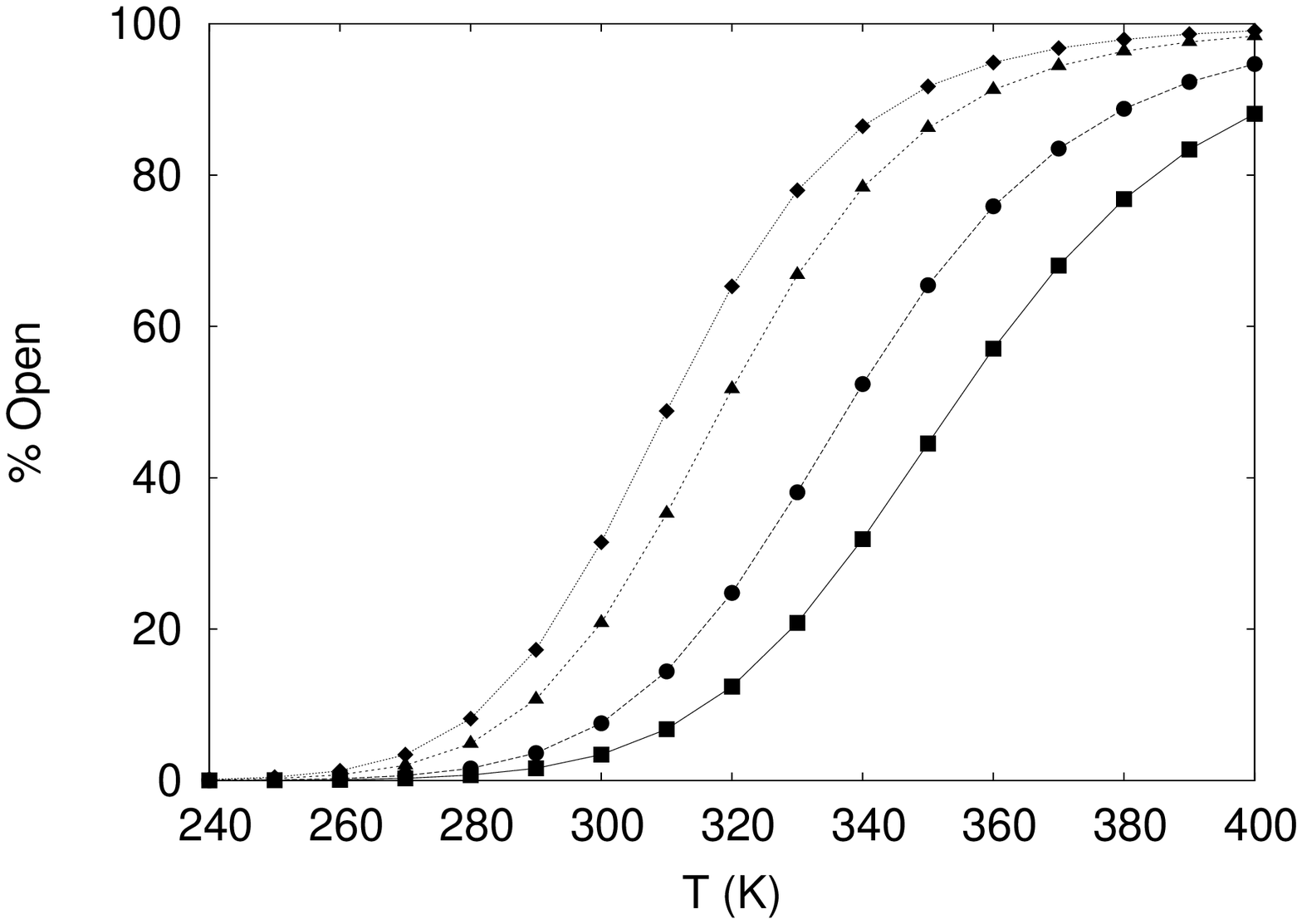} \\
$\epsilon=0.0012\;$eV/\AA$^2$ \\
\includegraphics[width=8cm]{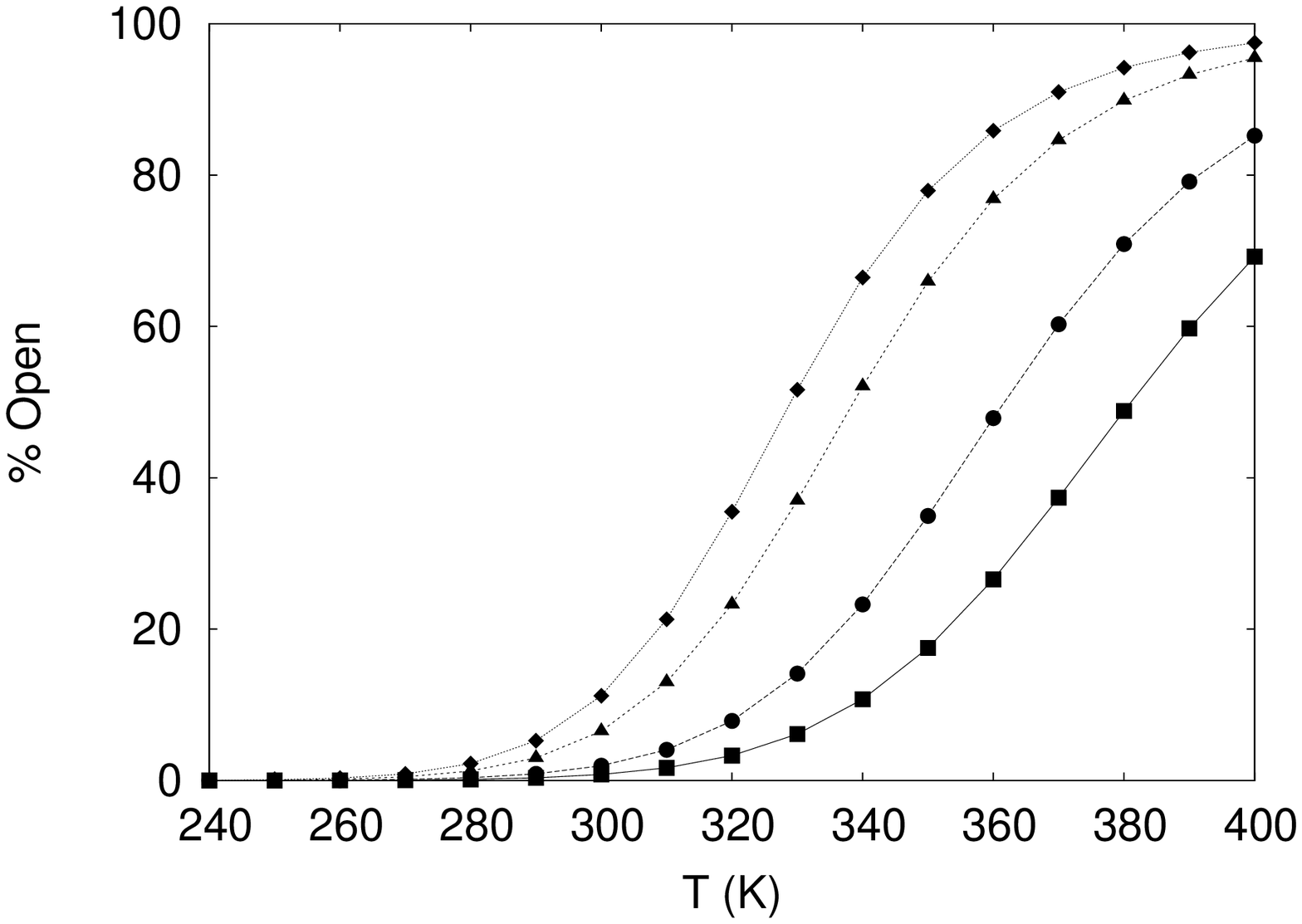} 
  \end{center}
\caption{Variation versus temperature of the percentage of open hairpins
for two values of $\epsilon$ and different loop lengths: $L=10$
(squares) $L=14$ (circles), $L=24$ (triangles) and $L=32$ (diamonds). 
}
\label{fig:meltingcurves}
\end{figure}

There are however two aspects on which the model quantitatively
disagrees with experiments.
First it gives a width of the melting transition which is
significantly larger than in experiments. The model finds that the
temperature range over which the percentage of open hairpins varies
from 20\% to 80\% extends above approximately $50\;$K (depending on
$L$) while experiments measure a range of about $15\;$K for poly(T)
loops and about $30\;$K for poly(A). Second, 
as shown in Fig.~\ref{fig:meltingcurves}, the model gives a
variation of $T_m$ versus $L$ which is approximately the same for
poly(A) ($\epsilon=0.0016\;$eV/\AA$^2$) and for poly(T)
($\epsilon=0.0012\;$eV/\AA$^2$), while experiments indicate that the
effect of the loop length $L$ should be significantly larger for
poly(A) than for poly(T).

\begin{figure}[h!]
  \begin{center}
$\epsilon=0.0016\;$eV/\AA$^2$ \\ 
\includegraphics[width=8cm]{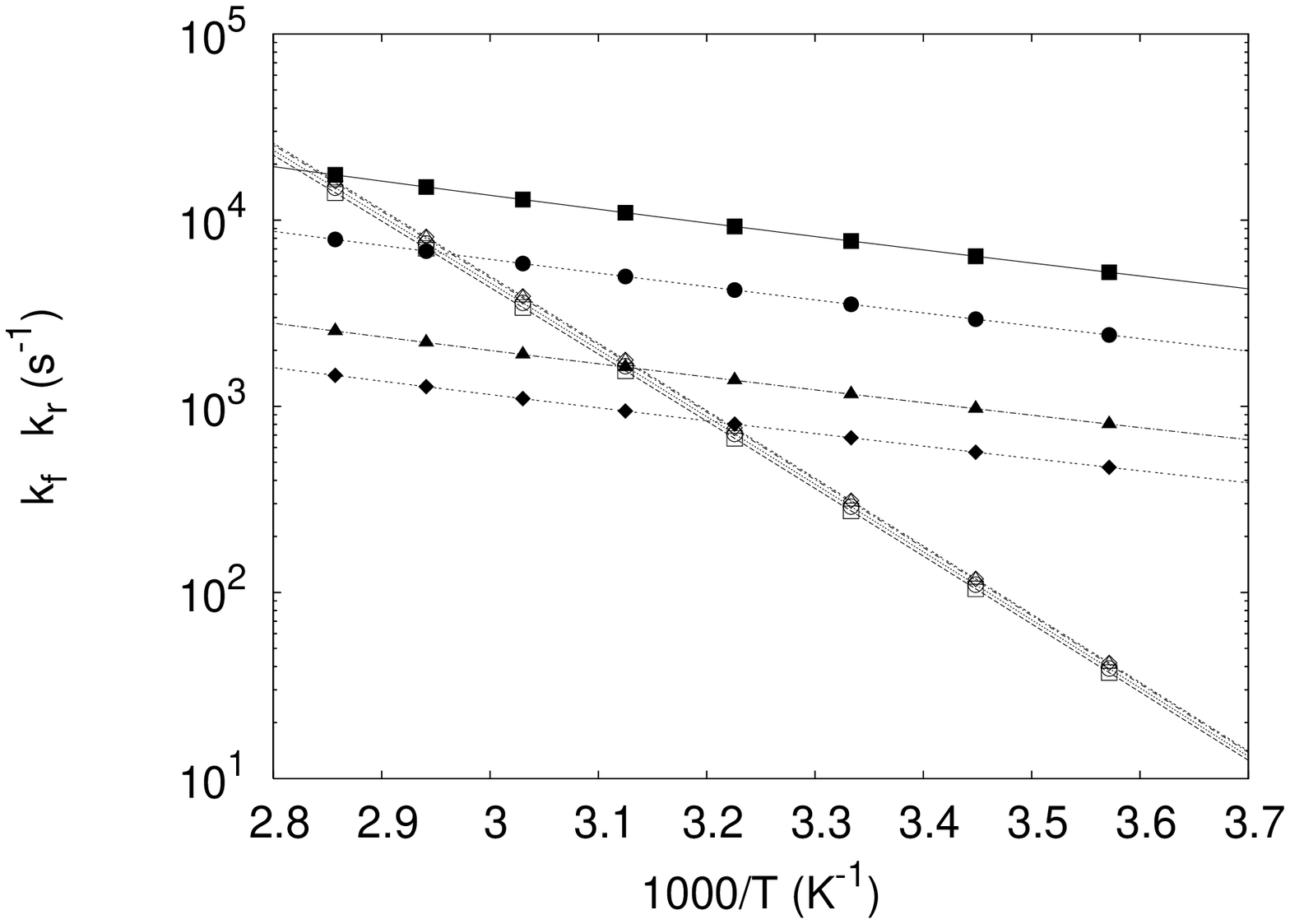} \\
$\epsilon=0.0012\;$eV/\AA$^2$ \\
\includegraphics[width=8cm]{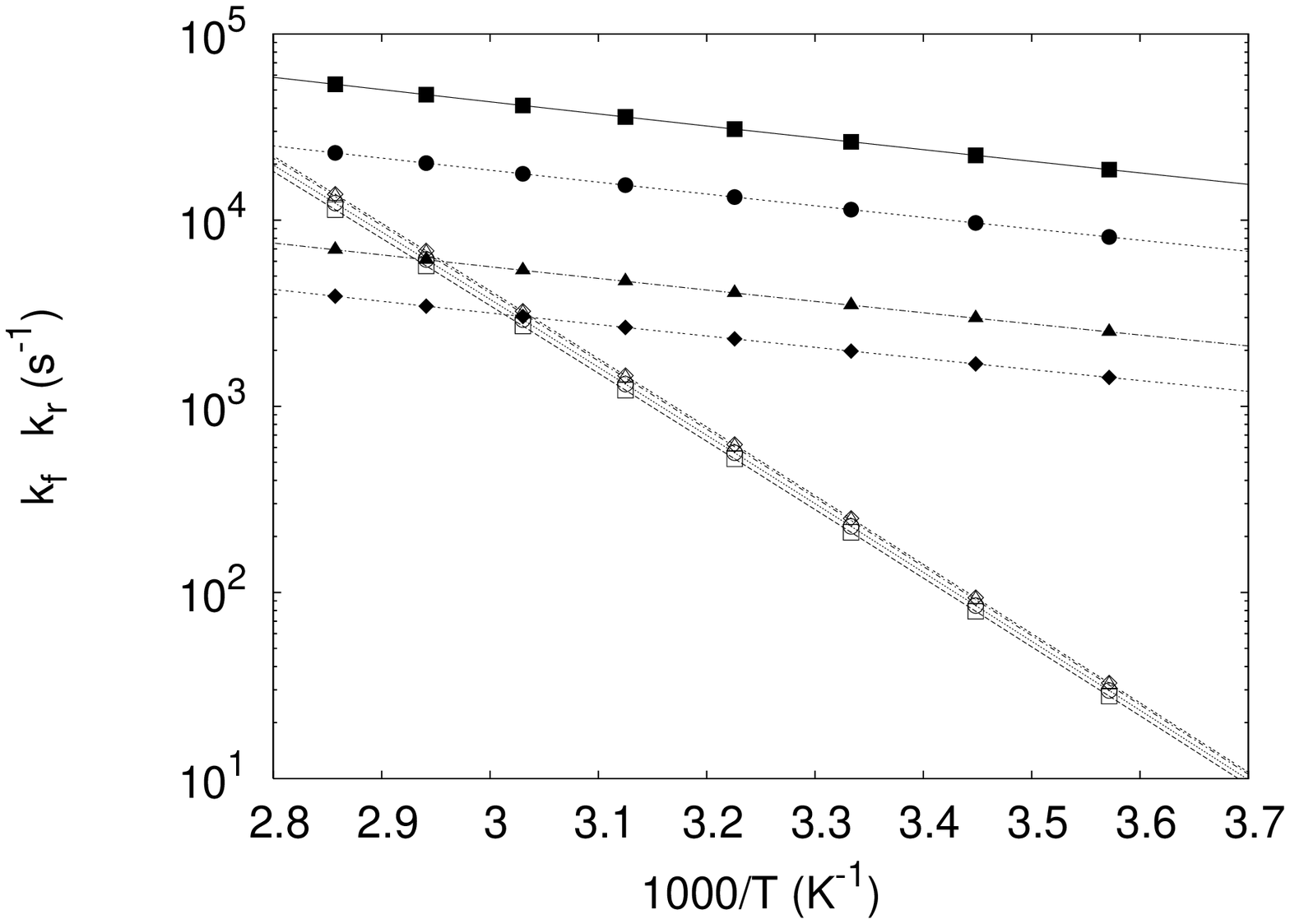} 
  \end{center}
\caption{Variation versus temperature of the reaction rates
for opening $k_f$ (open symbols) and for closing $k_r$ (closed symbols)
for two values of $\epsilon$ and different loop lengths: $L=10$
(squares) $L=14$ (circles), $L=24$ (triangles) and $L=32$ (diamonds). 
The reaction rates are plotted in
  logarithmic scale, versus $1000/T$. The calculations have been made
  with $D_0 = 3.0\; 10^6\;$cm$^2$/s.
}
\label{fig:kmulti}
\end{figure}

\begin{figure}[h!]
\includegraphics[width=8cm]{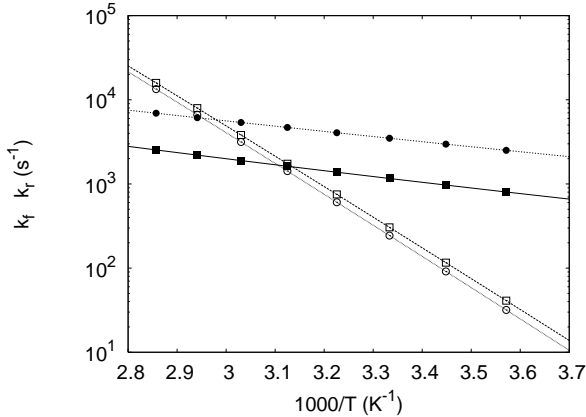}
\caption{Comparison of the temperature variation of 
the reaction rates
for opening $k_f$ (open symbols) and for closing $k_r$ (closed
symbols) for two values of $\epsilon$: 
squares $\epsilon=0.0016\;$eV/\AA$^2$ (poly(A)), circles
$\epsilon=0.0012\;$eV/\AA$^2$ (poly(T)).
}
\label{fig:kseq}
\end{figure}

\begin{figure}[h!]
\includegraphics[width=8cm]{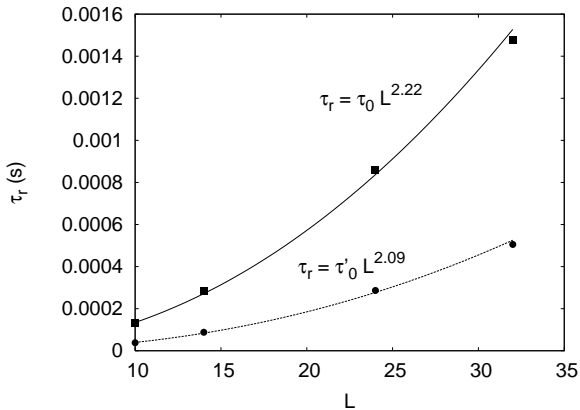}
\caption{Closing times of the hairpin at $300\;$K,
$\tau_r = k_r^{-1}$, versus $L$ for
two values of $\epsilon$: 
squares $\epsilon=0.0016\;$eV/\AA$^2$ (poly(A)), circles
$\epsilon=0.0012\;$eV/\AA$^2$ (poly(T)). The points are the numerical
values given by the model and the lines are fits according to the
formula indicated in the graph.
}
\label{fig:kversusL}
\end{figure}

\bigskip
Figure \ref{fig:kmulti} shows the variation versus $T$ of the reaction
rates for opening $k_f$ and closing $k_r$ for different loop lengths,
for two values of $\epsilon$ describing poly(A) and poly(T) loops. 
As noted
in subsection \ref{subsec:results}, the
order of magnitude of the values that we obtain for the reaction rates
are in agreement with the experimental results
\cite{BONNET98}. Another important point is that $k_f$ is 
nearly independent of the loop length (Fig.~\ref{fig:kmulti}) or loop
sequence (Fig.~\ref{fig:kseq}), as observed experimentally. 
The variation of $k_f$ versus $T$ is well described by an Arrhenius
law with an activation energy $E_o = 0.74\;$eV (or $17\;$kcal/mol,
while experiments report a higher value of $34\;$kcal/mol).
Conversely
the closing rate depends on the loop length or sequence. Lower rates
are obtained for longer, or more rigid, loops, as one would expect
qualitatively by considering that closing is mainly determined by the
random diffusion of the two sides of the loops that bind when they
find each other in space. It is interesting to examine the variation
of the closing time $\tau_r = k_r^{-1}$ 
versus the size of the loop $L$, shown in
Fig.~\ref{fig:kversusL}. It can be approximated by the scaling law
\begin{equation}
  \tau_r = \tau_0 \; L^{\phi}
\end{equation}
with an exponent $\phi = 2.09$ for the poly(T) case and $\phi = 2.22$
for the more rigid poly(A) case. These values should be compared with
the values $\phi_0 = 2$ for a Gaussian chain or
$\phi_1 = 1.8$ obtained for a flexible polymer with excluded
volume effects \cite{deGennes}. Our results that give a lower exponent
$\phi$ when $\epsilon$ is reduced are consistent with this theoretical
predictions. 
Some experimental results report an
exponent of $2 \pm 0.2$ \cite{Ansari}, but the scaling was measured on
very small loop ($4 \le L \le 12$). While the closing rate depends
strongly on the loop sequence, in the temperature range that we
investigated it is well described by an Arrhenius law with an
activation energy that depends weakly on the sequence. For
$L = 24$, we get $E_c = 0.148\;$eV ($3.4\;$kcal/mol) for 
 $\epsilon=0.0016\;$eV/\AA$^2$ (poly(A)) and $E_c = 0.128\;$eV
($2.96\;$kcal/mol) for $\epsilon=0.0012\;$eV/\AA$^2$ (poly(T)).

\bigskip
The results presented up to now have been obtained by describing
the DNA strands with a KP polymer model. This model is
interesting because it allows us to describe the energetic effects
associated to the bending of the strand. However, as we have seen that
the results exhibit some limitations of the hairpin model, it is interesting
to examine the influence of the model chosen to describe the
properties of the loop. Figures \ref{fig:probafitFRC} to
\ref{fig:kversusLFRC} show the results obtained if we consider the
strand as a Freely Rotating Chain (FRC) \cite{Flory}, i.e.\ a  
polymer made of segments of length $\ell$, such that two consecutive
segments make a fixed angle $\theta$ but can rotate freely around each
other (Fig.~\ref{fig:modelFRC}). The energy of a FRC chain
is a constant and the contribution of the polymer is only entropic.

\begin{figure}[h!]
\begin{center}
\includegraphics[width=4cm]{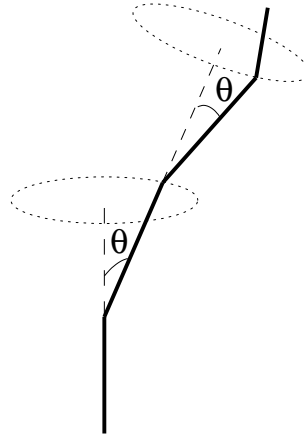}
\end{center}
\caption{The Freely Rotating Chain polymer model. The angle between
  consecutive segments is fixed, and each segment can rotate freely
  around the axis defined by the previous one.
}
\label{fig:modelFRC}
\end{figure}

\bigskip
The probability distribution function $P^{FRC}(\rho)$
of the FRC 
cannot be expressed analytically but it is easy to obtain it by a
Monte Carlo simulation. This numerical expression can be introduced in
the calculation of the constrained partition function according to
Eq.~(\ref{eq:Zrho1}), but, as for the KP chain, we need an
analytical expression of $S(\rho'|\rho)$ to carry out the calculations.
Figure \ref{fig:probafitFRC} shows that it can again be provided by an
effective Gaussian approximation determined by choosing the parameter
$\tau$ according to $L \tau^2 = \langle \rho^2 \rangle / 6$. For the
FRC, one has $\langle \rho^2 \rangle = L \; \ell^2
(1 + \cos \theta)/(1 - \cos \theta)$ so that the value of $\chi$ to be
entered in the expressions (\ref{eq:tau2}) and (\ref{eq:Sgauss1}) 
is $\chi = (1 + \cos \theta)/(1 - \cos \theta)$. 

In order to compare the two polymer models, we have selected for the
FRC case values of $\theta$ which give a persistence length
comparable to the cases that we investigated for the Kratky Porod
model. The matching cannot be perfect because, as the FRC model has a
constant energy, its persistence length does not depend on
temperature, contrary to the KP case. We have selected the
values of $\theta$ so that the persistence length of the two models
match at $T=300\;$K. For the FRC model, the persistence length is
\cite{Flory} 
\begin{equation}
  \label{eq:lambdaFRC}
\lambda' = - \dfrac{\ell}{\ln(\cos \theta)} \; .
\end{equation}
The values $\theta = 54.945\;^{\circ}$ and $\theta = 61.667\;^{\circ}$
give the same persistence lengths as the KP model at
$300\;$K for $\epsilon=0.0016\;$eV/\AA$^2$ (poly A) and
$\epsilon=0.0012\;$eV/\AA$^2$ (poly T).

\begin{figure}[h!]
\begin{center}
$L = 24 $\\
 \includegraphics[width=8cm]{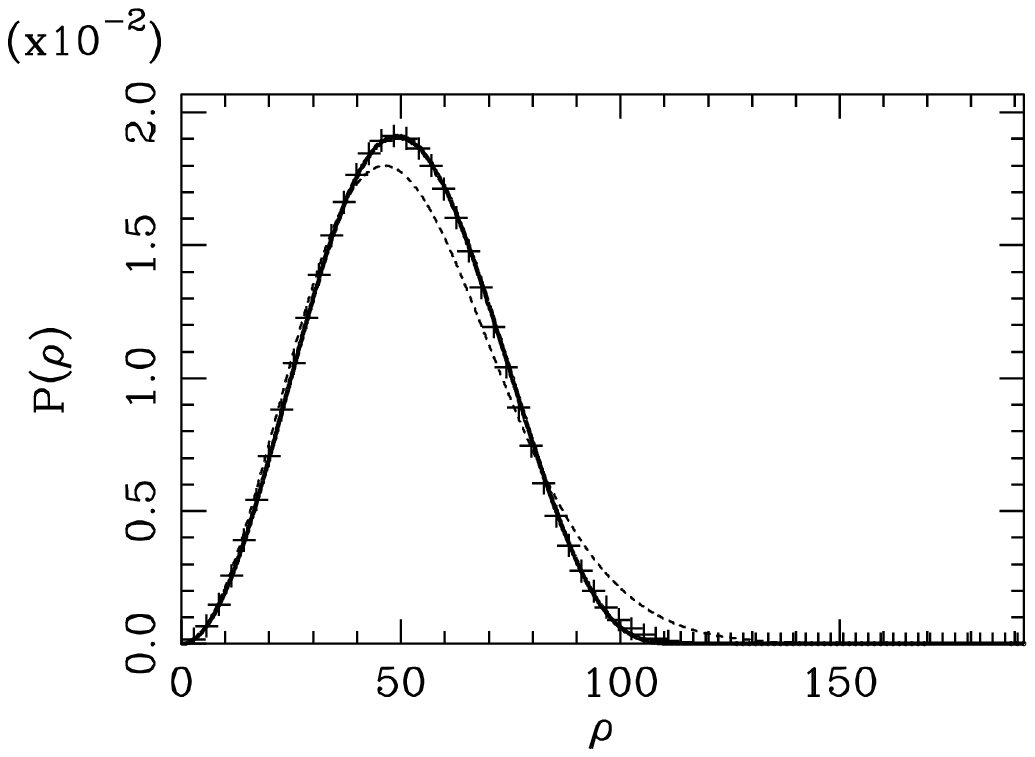} \\
 $L = 14$ \\
\includegraphics[width=8cm]{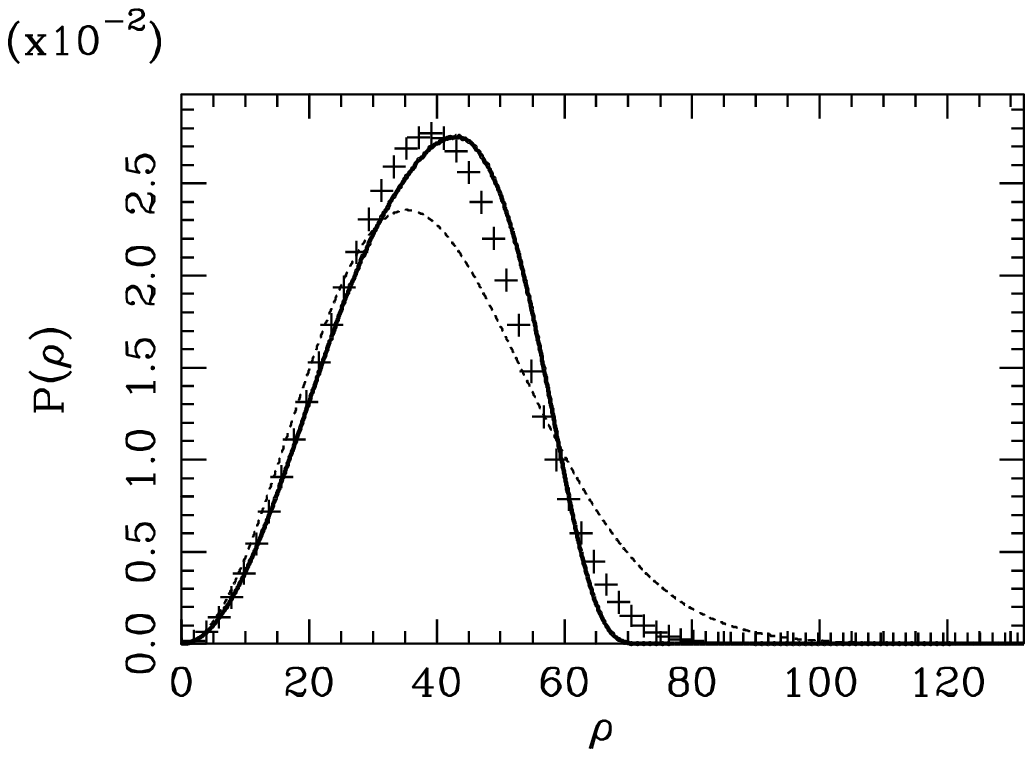}
\end{center}
\caption{Comparison between the FRC
  distribution function $P^{FRC}_L(\rho)$ (full line) and the Gaussian
  approximation $P^G_L(\rho)$ (dashed line) for two values of $L$. The crosses
  show the distribution function $P_L(\rho)$ obtained
  by starting from the FRC distribution $P^{FRC}_{L-2}(\rho)$
  and computing the probability distribution of the end-to-end
  distance of a polymer extended by two units, using the conditional
  probability $S(\rho'|\rho)$ according to Eq.~\ref{eq:PSP}.
The parameters for the FRC model are $\theta = 54.945\;^{\circ}$,
  $\ell = 6\;$\AA, giving a persistence length $\lambda = 10.8\;$\AA.
}
\label{fig:probafitFRC}
\end{figure}

\begin{figure}
  \includegraphics[width=8cm]{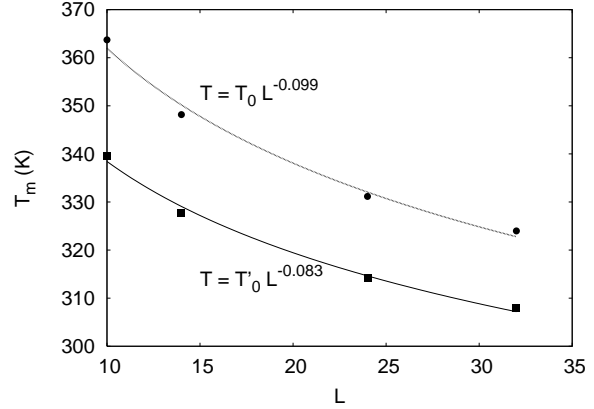}
\caption{Variation of the opening temperatures of DNA hairpins deduced
  from the model where the loop is described by a FRC model
  for $\theta = 54.945\;^{\circ}$ (circles) and
  $\theta = 61.667\;^{\circ}$ (squares). The curves show fits with
  the function $T = T_0 L^{\nu}$.
}
\label{fig:TmFRC}
\end{figure}

\begin{figure}[h!]
  \begin{center}
$\theta = 54.945\;^{\circ}$ \\
\includegraphics[width=8cm]{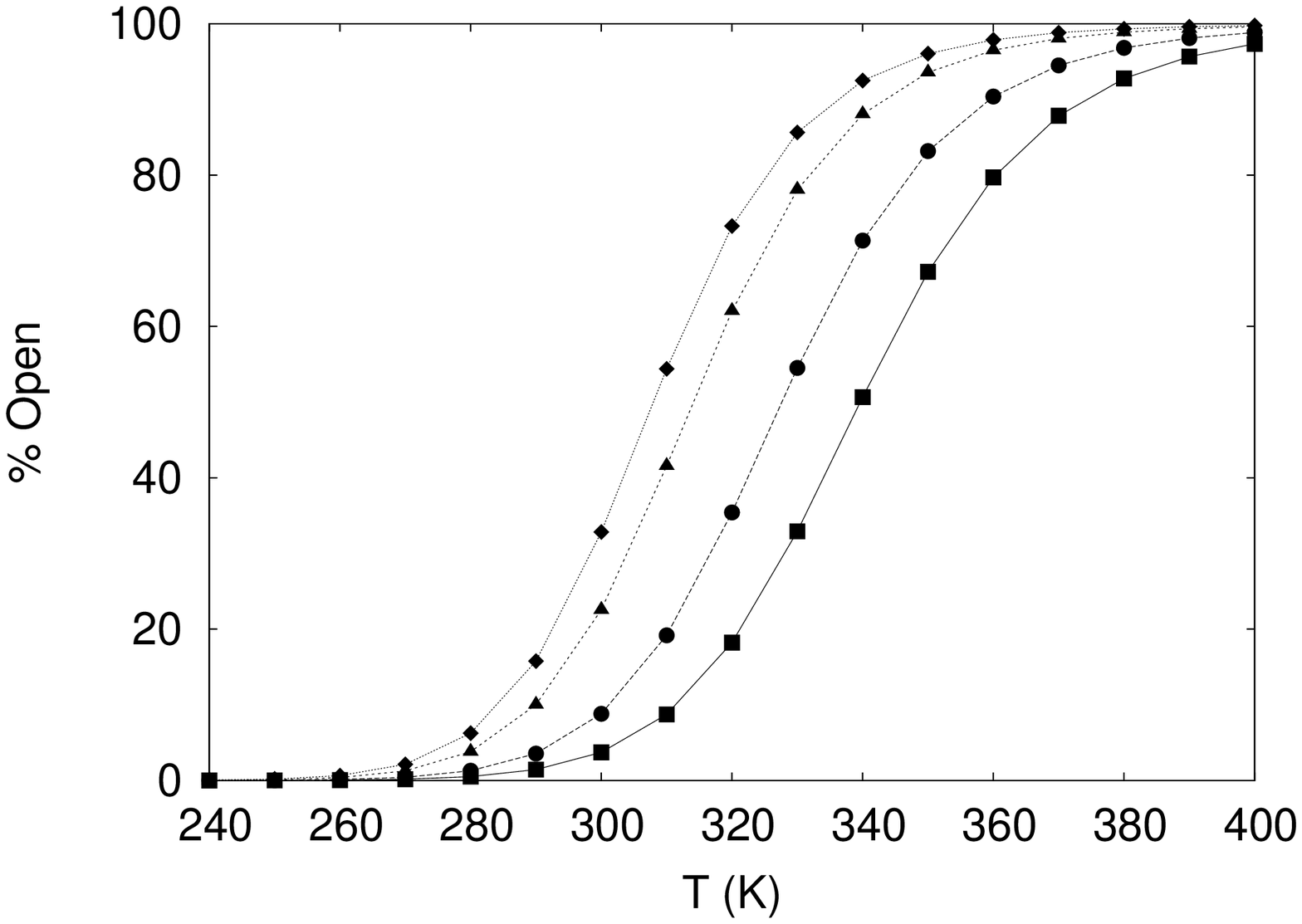} \\
$\theta = 61.667\;^{\circ}$ 
\includegraphics[width=8cm]{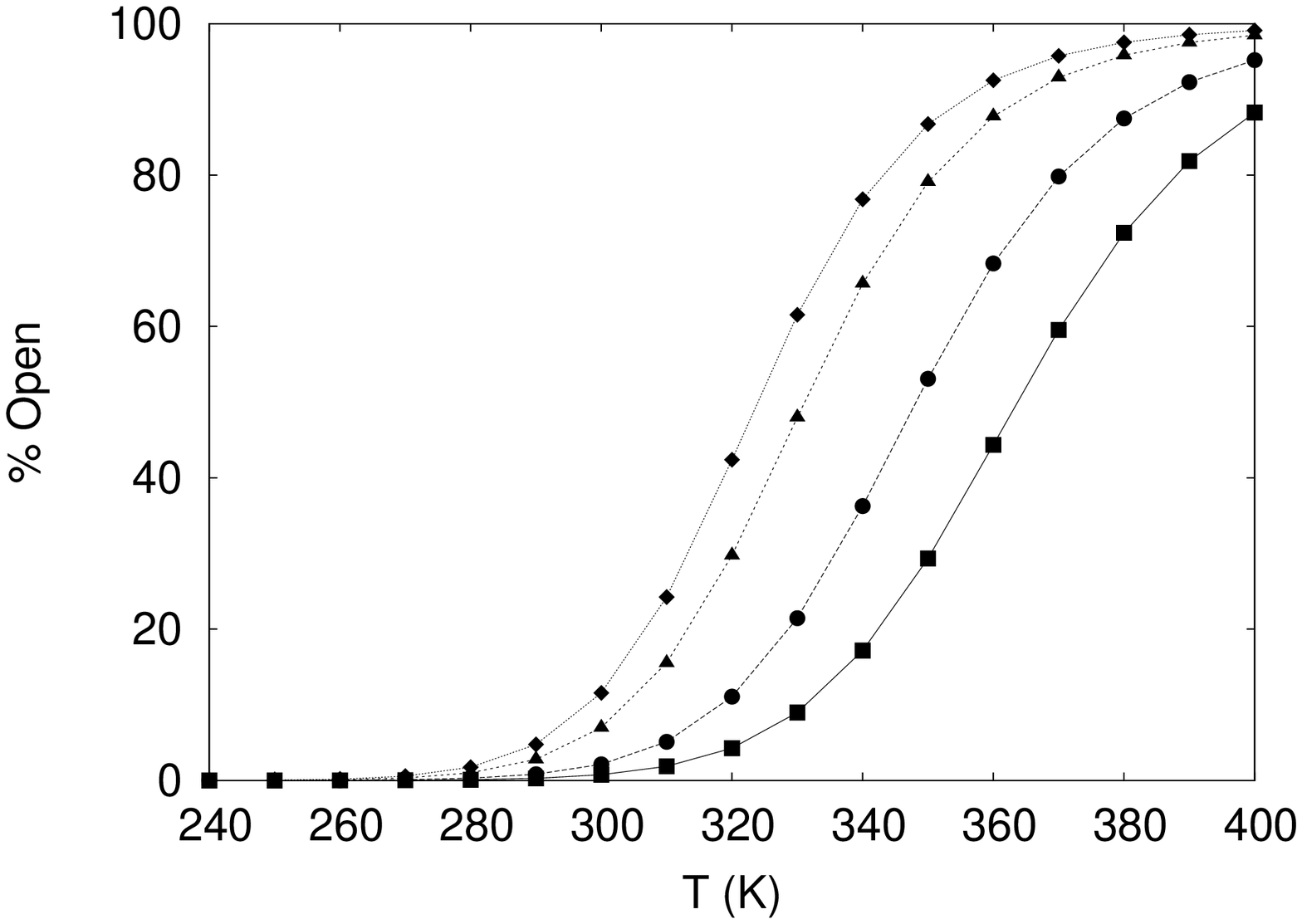} 
  \end{center}
\caption{Variation versus temperature of the percentage of open
  hairpins
with a FRC model of the loop
for two values of $\theta$ and different loop lengths: $L=10$
(squares) $L=14$ (circles), $L=24$ (triangles) and $L=32$ (diamonds). 
}
\label{fig:meltingcurvesFRC}
\end{figure}

\begin{figure}[h!]
  \begin{center}
$\theta = 54.945\;^{\circ}$ \\
\includegraphics[width=8cm]{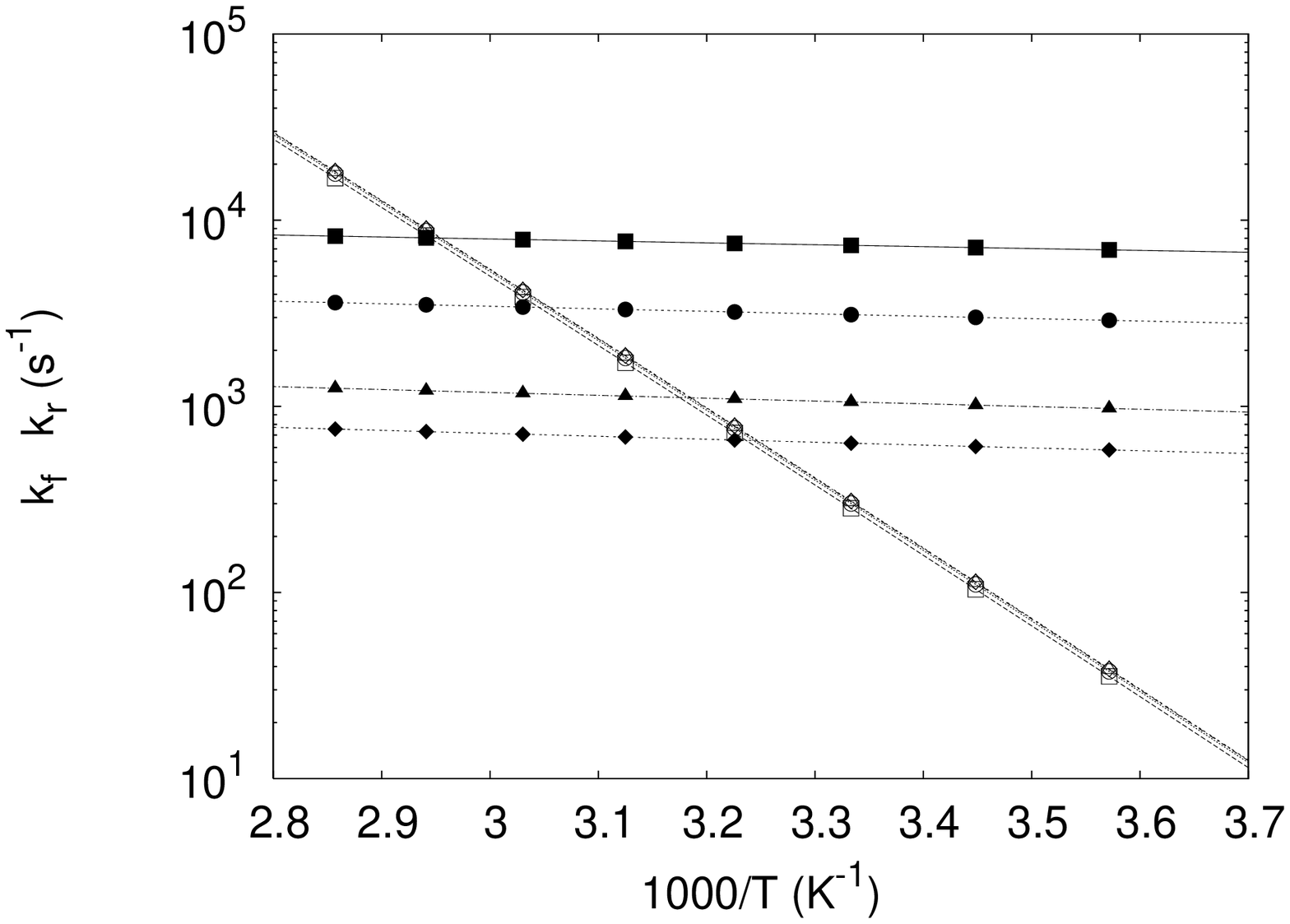} \\
$\theta = 61.667\;^{\circ}$ \\
\includegraphics[width=8cm]{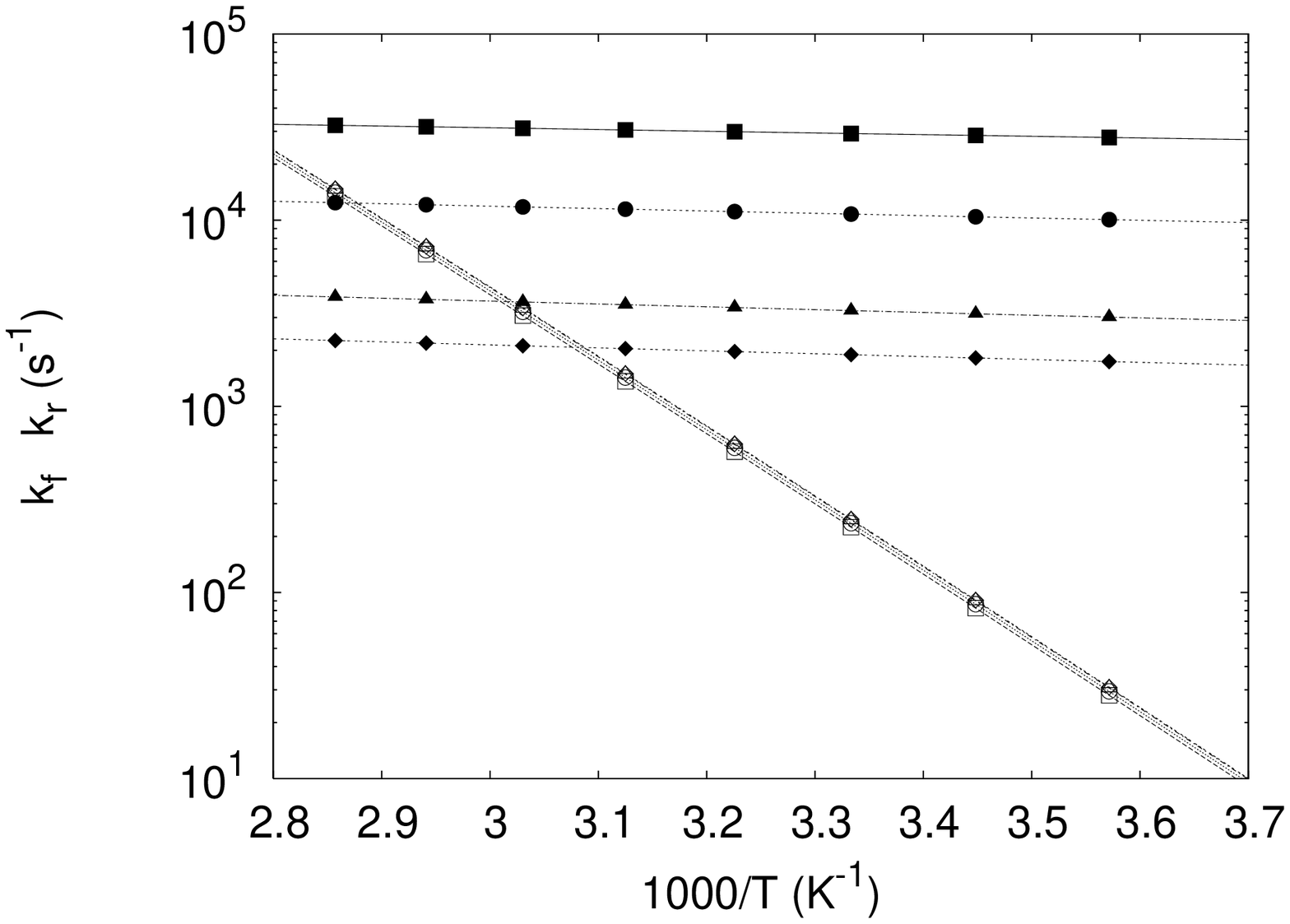} 
  \end{center}
\caption{Variation versus temperature of the reaction rates
for opening $k_f$ (open symbols) and for closing $k_r$ (closed
symbols)
with a FRC model of the loop
for two values of $\theta$ and different loop lengths: $L=10$
(squares) $L=14$ (circles), $L=24$ (triangles) and $L=32$ (diamonds). 
The reaction rates are plotted in
  logarithmic scale, versus $1000/T$. The calculations have been made
  with $D_0 = 3.0\; 10^6\;$cm$^2$/s.
}
\label{fig:kmultiFRC}
\end{figure}

\begin{figure}[h!]
\includegraphics[width=8cm]{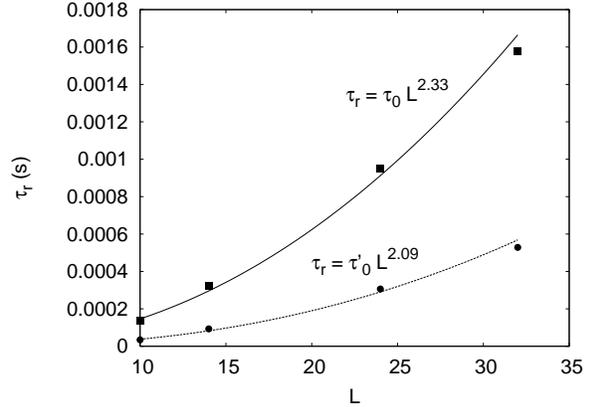}
\caption{Closing times of the hairpin at $300\;$K,
$\tau = k_r^{-1}$, versus $L$ for
a FRC model of the loop for two values of $\theta$: 
squares $\theta = 54.945\;^{\circ}$ (poly(A)), circles
$\theta = 61.667\;^{\circ}$ (poly(T)). The points are the numerical
values given by the model and the lines are fits according to the
formula indicated in the graph.
}
\label{fig:kversusLFRC}
\end{figure}

The comparison of figures~\ref{fig:TmFRC} to \ref{fig:kversusLFRC} for the
FRC model with the corresponding figures with the
KP  model shows that most of the results are qualitatively 
similar
for both models. The melting curves of
Fig.~\ref{fig:meltingcurvesFRC} for the FRC model exhibit a narrower
temperature range for melting than the corresponding curves of
Fig.~\ref{fig:meltingcurves} for the KP model, which would
be closer to experimental observations. But both models show a larger
variation of $T_m$ when the persistence length of the loop is larger,
which disagrees with the observations.

Figures \ref{fig:kmultiFRC} for the FRC model and \ref{fig:kmulti} in
the KP case show the same general behavior that the opening
rate $k_f$ is almost independent of the length of the loop, 
whereas
the closing rate $k_r$ varies by more than one order of magnitude when
$L$ changes from 10 to 32. But there is a qualitative difference
between the FRC and Kratky Porod model (which is partly hidden by the
logarithmic scales of the figures) concerning the activation energy
for closing. While it was of the order of $0.13\;$eV (3 kcal/mol) for
the Kratky Porod model, in agreement with experiments, it is 5 times
smaller for the FRC model ($\approx 0.57\;$kcal/mol). This is
consistent with the absence of any energy contribution in the FRC,
whose rigidity is described in purely geometrical terms, and in fact
points out the model's deficiency in describing the properties of DNA
strands.

The effect of the size of the loop on the closing times of the
hairpins at $300\;$K is very similar for the FRC and the KP
models (Figs.~\ref{fig:kversusLFRC} and \ref{fig:kversusL}) 
because the parameters of the two models have been selected to
give the same persistence lengths at this temperature.

\bigskip
In summary, the comparison between alternative descriptions of the
polymer properties of DNA single strands shows that different models
can bring some quantitative differences but that the qualitative
results are not changed; the main discrepancy between theory and
experiments concerning the variation of $T_m$ versus $L$ for different
loop lengths, which is greater for more rigid loops in experiments
while the theory gives the opposite, does not seem to be resolved
simply by using another polymer model.

\section{Conclusion}

In this study we presented a theoretical model of the physics of DNA
hairpin formation and melting which tries to capture the essential
phenomena within a highly simplified picture. Basically it combines a
model for the double helix assembly with standard polymer concepts. This
approach exhibits successes and weaknesses which are themselves
instructive for understanding the properties of DNA and RNA strands.

Our mesoscopic approach provides acceptable systematics for
thermodynamic and kinetic properties of hairpins with a poly(T)
loop. Using realistic parameters for the binding energies, persistence
length of the loop and diffusion coefficient of the polymer, the 
KP variant of the model
describes the variation of $T_m$ versus $L$ and the
order of magnitude of opening and closing times. It shows that the
kinetics of the opening is almost unaffected by the length of the
loop, in agreement with experiments. Closing times increase very
significantly for longer loops while the corresponding activation
energy is almost independent of $L$, as observed experimentally. The
enthalpy of closing is quantitatively described while the enthalpy of
opening given by the model is only half of the observed value. This
aspect is related to the temperature range over which the melting
transition is found in the model, which is significantly broader than
in the experiments. Although the model is only semi-quantitative in
some respects, it is nevertheless able to describe a whole set of
equilibrium and non-equilibrium data with a small set of realistic
parameters. It should be stressed that studying thermodynamics {\em
  and} kinetics in the same framework is a rather demanding test.

\smallskip
Weaknesses appear when one tries to apply the theory to poly(A)
loops. The model correctly detects that $T_m$ is lowered but it finds
that the variation of $T_m$ with $L$ is smaller for poly(A) than for
poly(T). Intuitively this makes sense because one can understand the
decrease of $T_m$ with $L$ as an entropic effect due to the
fluctuations of the loop. As poly(A), with its larger bases is
considered to be more rigid than poly(T) \cite{GODDARD}, which is
reflected in the higher value of $\epsilon$ that we introduce in the
KP description of poly(A), one can expect that this extra rigidity
reduces fluctuations thereby decreasing the entropy gain due to loop
extension. Thus basic physics leads to the conclusion that the effect
of the loop length should be smaller for poly(A) than for poly(T) but
experiments show exactly the contrary. Another discrepancy between
our model and experimental data appears when one examines the enthalpy
for closing given by the kinetic studies. Experiments find that
$\Delta H_c$ is approximately 5 times larger for poly(A) than for
poly(T), while we only get a small increase when $\epsilon$ is changed from
$0.0012\;$eV/\AA$^2$ to $0.0016\;$eV/\AA$^2$. Varying parameters one
can increase $\Delta H_c$ for poly(A) in the model, but the disagreement
with experiments is transfered elsewhere, in particular on
$T_m$. These discrepancies between theory and experiments for poly(A)
are frustrating but probably also very instructive. They suggest that
``rigidity'' is not the only feature that distinguishes poly(A) from
poly(T), otherwise the KP model would be able to describe it. It
appears that the effect of having large bases which can stack on each
other is deeper and might not be captured by a simple polymer chain
model.

\smallskip
In conclusion, attempting to put the
thermodynamic and kinetic properties of DNA hairpins 
in the same model framework remains a
challenge. Our results show that the role of the loop is decisive
and, for poly(A), extends beyond a simple rigidity effect. This indicates
that experiments on hairpins are very sensitive probes of the
properties of single-stranded DNA on a scale of a few tens of base
pairs. In other words, beacons tell us not only about themselves but
also about mesoscopic properties of single-stranded DNA and RNA, which
have a high biological relevance.

\appendix
\section{Calculation of the conditional probability $S(\rho'|\rho)$
  for a Gaussian chain.}
\label{app:srhorhop}

Let us consider a Gaussian chain made of orientationally
uncorrelated links such that the probability for any segment to lie
along a vector $\vec{\Delta}$ is proportional to \par \noindent $\exp(-
|\vec{\Delta}|^2 / 4 \tau^2 )$. The probability that the end-to-end
distance of a chain of $L$ monomers is at distance $\rho$ is then
\begin{equation}
  \label{eq:pgauss}
P^G(\rho) = \dfrac{1}{2 \sqrt{\pi }} \dfrac{1}{\sigma}  \left( 
\dfrac{\rho}{\sigma} \right)^2 e^{- \rho^2 / 4 \sigma^2} \; ,
\end{equation}
with $\sigma^2 = L \tau^2$, where the $\rho^2$ prefactor comes from
the integration over all orientations of the end-to-end vector. It is
such that $\langle \rho^2 \rangle =L\; \ell^2 = 6 \sigma^2$. Consider
such a Gaussian chain with an end-to-end vector
$\bm {\rho}$, 
and assume that we add to each end segments $\vec{\Delta}_1$ and
  $\vec{\Delta}_2$. Its end-to-end vector becomes 
$\bm{\rho}\,' = \bm{\rho} +
\vec{\Delta}_1 - \vec{\Delta}_2$ and
the conditional probability that the end-to-end distance of the
extended chain is $\rho'$, given $\rho$, is 
\begin{align}
  S(\rho'|\rho) = \;& A  \rho'^2 \int d\Omega_{\rho'} \;
  \int d\vec{\Delta}_1 \; d\vec{\Delta}_2 \;
  \nonumber \\
& 
e^{ - (\Delta_1^2 + \Delta_2^2)/4\tau^2} \delta(\bm{\rho}\,'
- \bm{\rho} - \vec{\Delta}_1
+ \vec{\Delta}_2)
\end{align}
where $A$ is a normalization constant to be determined at the end of
the calculation, and where the first integral over the orientations of
$\bm{\rho}\,'$ is introduced because we are only interested in the
end-to-end distance of the chain. The integration over
$\vec{\Delta}_2$ is immediate. Let us define $\vec{u} = \bm{\rho}\,'
- \bm{\rho}$. Up to a normalization factor we get
\begin{align}
  S(\rho'|\rho) = \; & A \rho'^2 \int  d\Omega_{\rho'} 
\int d\vec{\Delta}_1 e^{-\Delta_1^2/4 \tau^2} \nonumber \\
&\int_{-1}^{+1} d\mu \; e^{-(\Delta_1^2 + u^é - 2 u \Delta_1 \mu)/4
  \tau^2} \;
\end{align}
where the integral over
\begin{equation}
  \mu = \dfrac{\vec{u} \cdot \vec{\Delta}_1}%
{u \; \Delta_1} 
\end{equation}
comes is the integration over the azimuthal angle of
$\vec{\Delta}_1$. This leads to
\begin{align}
S(\rho'|\rho) = A \rho'^2 &\int  d\Omega_{\rho'} \dfrac{1}{u} e^{-u^2/4
  \tau^2} \nonumber \\
&\int_0^{\infty} d\Delta_1 \; \Delta_1 e^{- \Delta_1^2/4
  \tau^2}
\sinh\left( \dfrac{u \Delta_1}{2 \tau_2} \right)
\end{align}
up to normalization factors. Using the definite integral
\begin{equation}
  J(a,b) = \int_0^{\infty} dx \; x \; e^{-a x^2} \; \sinh bx = 
\dfrac{b}{4a} \sqrt{\dfrac{\pi}{a}} e^{b^2/4a} \; ,
\end{equation}
we can perform the integration over $\Delta_1$. Reintroducing 
$\vec{u} = \bm{\rho}\,' - \bm{\rho}$,
and defining
\begin{equation}
  \eta = \dfrac{ \bm{\rho}\,' \cdot \bm{\rho}}%
{\rho' \; \rho} \; ,
\end{equation}
we get 
\begin{equation}
  S(\rho'|\rho) = A \rho'^2 \int_{-1}^{+1} d\eta \; 
e^{-(\rho'^2 + \rho^2 - 2 \rho' \rho \eta)/8 \tau^2} \;.
\end{equation}
Performing the final integration, and determining the normalization
constant from
\begin{equation}
  \int_0^{\infty} d\rho' S(\rho' | \rho) = 1 \quad \forall \rho \; ,
\end{equation}
we obtain
\begin{equation}
\label{eq:Sgauss1app}
 S(\rho'|\rho) = \sqrt{\dfrac{1}{2 \pi \tau^2}} \; \dfrac{\rho'}{\rho}
 e^{-(\rho'^2 + \rho^2)/8 
 \tau^2}
\sinh\left( \dfrac{\rho' \rho}{4 \tau^2} \right) \; .
\end{equation}

\section{Calculation of the first passage time in a diffusion
  controlled process.}
\label{app:calcultau}

We consider the Smoluchowski equation (\ref{eq:FP2}) for the
probability distribution ${\cal P}(\rho,t)$, with a 
function ${\cal F}(\rho)$ which
has the shape of a double well with a local maximum at
$\rho=\rho^{\star}$. Initially the system is assumed to be in the well
$\rho_0 < \rho < \rho^{\star}$ and we assume a reflecting condition
at the boundary $\rho = \rho_0$, which implies 
$j(\rho_0,t) = 0 \;\; \forall t$.
To determine the first passage time above the maximum at
$\rho^{\star}$ an absorbing boundary condition is assumed for $\rho =
\rho^{\star}$. It can be expressed as $j(\rho^{\star},t) = \kappa
{\cal P}(\rho^{\star},t)$ and taking the limit $\kappa \to \infty$.

The probability that the system is still in the original well at time
$t$ is $\int_{\rho_0}^{\rho^{\star}} {\cal P}(\rho,t) d\rho$, so that the
first passage time above the barrier is 
\begin{equation}
\label{eq:B1}
  \tau = \int_0^{\infty} dt \int_{\rho_0}^{\rho^{\star}} d\rho
  \> {\cal P}(\rho,t) 
  \; .
\end{equation}
and has been calculated in References \cite{Szabo}  and \cite{Deutsch}.
For the sake of completeness we give an outline of the 
derivation \cite{Deutsch} in the context of the present study.

Integrating Eq.~(\ref{eq:FP1}) with respect to $\rho$
we get an expression of $j(\rho,t)$,
which can be used to express the boundary condition at $\rho^{\star}$
as
\begin{equation}
\label{eq:B2}
  j(\rho^{\star},t) =  \kappa {\cal P}(\rho^{\star},t) = -
  \int_{\rho_0}^{\rho^{\star}} d\rho \dfrac{\partial
  {\cal P}(\rho,t)}{\partial t} \;,
\end{equation}
and write  Eq.~(\ref{eq:FP2}) as
\begin{align}
   \int_{\rho_0}^{r} d\rho \dfrac{\partial
  {\cal P}(\rho,t)}{\partial t} &=  D_0 \left[
\dfrac{\partial {\cal P}}{\partial \rho} + \beta \dfrac{\partial {\cal F}}{\partial
  \rho} {\cal P}
\right] \nonumber \\
&= D_0 \; e^{- \beta {\cal F}} \dfrac{\partial}{\partial r} \left[ e^{\beta
  {\cal F}} {\cal P} 
  \right]
\end{align}
Integrating over $r$ in the range $(R,\rho^{\star})$ we get
\begin{align}
  e^{ \beta {\cal F}(\rho^{\star})} {\cal P}(\rho^{\star},t) - 
 e^{ \beta {\cal F}(R)} {\cal P}(R,t) \nonumber \\
=
\int_R^{\rho^{\star}} \dfrac{dr}{D_0 \, e^{- \beta {\cal F}(r)}}
\int_{\rho_0}^r d\rho \dfrac{\partial {\cal P}(\rho,t)}{\partial t} \; .
\end{align}
Using the boundary condition (\ref{eq:B2}) we obtain ${\cal P}(R,t)$ as
\begin{align}
  {\cal P}(R,t) = - \dfrac{1}{\kappa} \dfrac{e^{-\beta {\cal F}(R)}}
{e^{-\beta {\cal F}(\rho^{\star})}} \int_{\rho_0}^{\rho^{\star}} d\rho
\dfrac{\partial {\cal P}}{\partial t}  \; \; - \nonumber \\
e^{-\beta {\cal F}(R)} \int_R^{\rho^{\star}} \dfrac{dr}{D_0 \, e^{-
    \beta {\cal F}(r)}} 
\int_{\rho_0}^r d\rho \dfrac{\partial {\cal P}(\rho,t)}{\partial t} 
\end{align}

Let us define $p_0(r)$ by
\begin{equation}
  p_0(r) = \dfrac{e^{-\beta {\cal F}(r)}}{\int_{\rho_0}^{\rho^{\star}} d\rho
e^{-\beta {\cal F}(\rho)}} \; ,
\end{equation}
which is the probability that the system is at position $r$ in the
first well, weighted in this well so that it
verifies $\int_{\rho_0}^{\rho^{\star}} p_0(r) dr = 1$. It leads to
\begin{eqnarray}
  {\cal P}(R,t) &=& - \dfrac{1}{\kappa} \dfrac{p_0(R)}{p_0(\rho^{\star})}
\int_{\rho_0}^{\rho^{\star}} d\rho
\dfrac{\partial {\cal P}}{\partial t}  \; \;   \\ &-&
p_0(R) \int_R^{\rho^{\star}} \dfrac{dr}{D_0 \, p_0(r)}
\int_{\rho_0}^r d\rho \dfrac{\partial {\cal P}(\rho,t)}{\partial t} 
\nonumber
\end{eqnarray}
Using this expression to calculate $\tau$ according to
Eq.~(\ref{eq:B1}) gives
\begin{eqnarray}
  \tau & = &\dfrac{1}{\kappa \, p_0(\rho^{\star})}
\int_{\rho_0}^{\rho^{\star}} d\rho \; {\cal P}(\rho,t=0) + \nonumber \\
&  &
\int_{\rho_0}^{\rho^{\star}} dR \; p_0(R) 
\int_R^{\rho^{\star}} \dfrac{dr}{D_0 \, p_0(r)}
\int_{\rho_0}^r d\rho  \, {\cal P}(\rho,t=0) \quad ,  \nonumber
\end{eqnarray}
where we used $lim_{t\to\infty} {\cal P}(\rho,t) = 0 \;\; \forall \rho$.
Now since the system is assumed to be at equilibrium in the well 
$\rho_0 < \rho < \rho^{\star}$ at $t=0$, it follows from the definition
of $p_0(\rho)$ that ${\cal P}(\rho, t=0)= p_0(\rho)$. Therefore
\begin{align}
  \tau = \dfrac{1}{\kappa \, p_0(\rho^{\star})} + 
\int_{\rho_0}^{\rho^{\star}} dR \; p_0(R) 
\int_R^{\rho^{\star}} \dfrac{dr}{D_0 \, p_0(r)}
\int_{\rho_0}^r d\rho \; p_0(\rho) \; ,
\end{align}
Let us define $H(r)$ by
\begin{equation}
\label{eq:B10}
  H(r) = \dfrac{1}{D_0 \, p_0(r)}
\int_{\rho_0}^r d\rho \; p_0(\rho) \; .
\end{equation}
We have
\begin{align}
  \tau &=  \dfrac{1}{\kappa \, p_0(\rho^{\star})} +
\int_{\rho_0}^{\rho^{\star}} dR \; p_0(R)
\int_{\rho_0}^{\rho^{\star}}  dr \; H(r)  \; \Theta(r-R)
\nonumber \\
 &= \dfrac{1}{\kappa \, p_0(\rho^{\star})} +
\int_{\rho_0}^{\rho^{\star}} dr \; H(r) 
\int_{\rho_0}^{\rho^{\star}} dR \; p_0(R) \; \Theta(r-R) 
\nonumber \\
 &= \dfrac{1}{\kappa \, p_0(\rho^{\star})} +
\int_{\rho_0}^{\rho^{\star}} dr \; H(r) 
\int_{\rho_0}^{r} dR \; p_0(R)
\end{align}
where $\Theta(x)$ is the Heaviside step function. If we replace $H(r)$
by its expression (\ref{eq:B10}), we obtain
\begin{align}
  \tau &=  \dfrac{1}{\kappa \, p_0(\rho^{\star})} +
\int_{\rho_0}^{\rho^{\star}} dr \dfrac{1}{D_0 \, p_0(r)}
\left[\int_{\rho_0}^{r} dR \; p_0(R)\right]^2
\end{align}
Taking the limit $\kappa \to \infty$ corresponding to the absorbing
boundary condition when the system escapes above the barrier, we
finally obtain
\begin{equation}
  \tau = \int_{\rho_0}^{\rho^{\star}} dr \dfrac{1}{D_0 \, p_0(r)}
  I^2(r) \; ,
\end{equation}
with
\begin{equation}
  I(r) = \int_{\rho_0}^{r} dR \; p_0(R) \; ,
\end{equation}
which is the result of Eqs.~(\ref{eq:resutau}) and (\ref{eq:Ir}).


\begin{thebibliography}{xx}

\bibitem{BONNET99b}
G. Bonnet and A. Libchaber,
Physica A {\bf 263}, 68
(1999).

\bibitem{PENG}
Xiang-Hong Peng, Ze-Hong Cao, Jin-Tang Xia, G.W. Carlson, M.M. Lewis,
W.C. Wood, and L. Yang,
Cancer Res. {\bf 65}, 1909
(2005).

\bibitem{SANTANGELO}
P.J. Santangello, B. Nix, A. Tsourkas and G. Bao,
Nucleic Acid Research {\bf 32}, e57 (2004).

\bibitem{TAKINOUE}
M. Takinoue and A. Suyama,
Chem-Bio Informatics Journal {\bf 4}, 93
(2004).

\bibitem{SAKAMOTO}
K. Sakamoto, H. Gouzu, K. Komiya, D. Kiga, S. Yokoyama, T. Yokomori
and M. Hagiya,
Science {\bf 288}, 1223
(2000).

\bibitem{BONNET98}
G. Bonnet, O. Krichevsky and A. Libchaber,
Proc. Natl. Acad. Sci. USA, {\bf 95}, 8602
(1998).

\bibitem{KUZNETSOV}
S.V. Kuznetsov, Y. Shen, A.S. Benight and A. Ansari,
Biophysical J. {\bf 81}, 2864
(2001).


\bibitem{KratkyPorod}
O. Kratky and G. Porod,
Recl. Trav. Chim Pays Bas {\bf 68}, 1106
(1949).

\bibitem{PB}
 M. Peyrard, A.R. Bishop 
Phys. Rev. Lett. {\bf 62}, 2755
(1989).

\bibitem{DPB}
T. Dauxois, M. Peyrard and A.R. Bishop 
Phys. Rev. E {\bf 47}, 684
(1993).

\bibitem{differencePB}
Note that the stacking interaction that we use here is different from
the expression used in the PBD model 
$\frac{1}{2} K \Big[ 1 + \rho \exp[ - \zeta (y_m
    + y_{m-1})] \Big] \; \big(y_m - y_{m-1}\big)^2 $ because the PBD
model does not include an explicit description of the strands,
and must not allow a complete vanishing of the interaction, which
would mean a breaking of the DNA strand.

\bibitem{WilhelmFrey}
J. Wilhelm and E. Frey,
Phys. Rev. Lett. {\bf 77}, 2581
(1996).

\bibitem{HAMPRECHT}
B. Hamprecht and H. Kleinert,
Phys. Rev. E {\bf 71}, 031803
(2005).

\bibitem{SamSinh}
J. Samuel and S. Sinha,
Phys. Rev. E {\bf 66}, 050801 (2002).

\bibitem{StepSch}
S. Stephanow and G.M. Sch\"utz,
Europhys. Lett. {\bf 60}, 546 (2002).

\bibitem{Fisher}
M.E. Fisher,
Am. J. Phys. {\bf 32}, 343
(1964).

\bibitem{Marko}
This result has been derived in a slightly different form - which
includes the case of an external force - by 
J. Yan, R. Kawamura and J. Marko, 
Phys. Rev. E  {\bf 71}, 061905 (2005). 

\bibitem{Schulten}
K. Schulten, Z. Schulten, and A. Szabo, 
J. Chem. Phys. {\bf 74}, 4426
(1981).

\bibitem{Szabo}
A. Szabo, K. Schulten and Z. Schulten,
J. Chem. Phys. {\bf 72}, 4350
(1980).

\bibitem{Deutsch} 
J.M. Deutsch, 
J. Chem. Phys. {\bf 73}, 4700
(1980).

\bibitem{Stellwagen}
E. Stellwagen and N.C. Stellwagen,
Electrophoresis {\bf 23}, 2794
(2002).

\bibitem{GODDARD}
N.L. Goddard, G. Bonnet, O. Krichevsky and A. Libchaber,
Phys. Rev. Lett. {\bf 85}, 2400
(2000).

\bibitem{Smith}
S.B. Smith, Y. Cui and C. Bustamante,
Science {\bf 271}, 795
(1996).

\bibitem{Rivetti}
C. Rivetti, C. Walker and C. Bustamante,
J. Mol. Biol. {\bf 280}, 41
(1998).

\bibitem{Cuesta}
S. Cuesta L\'opez and Y.H. Sanejouand, private communication.

\bibitem{deGennes}
P.G. de Gennes, {\it Scaling concepts in Polymer Physics.}
Cornell University Press, N.Y., 1979.

\bibitem{Ansari}
A. Ansari, Y. Shen and S.V. Kuznetsov,
Phys. Rev. Lett. {\bf 88}, 069801
(2002).

\bibitem{Flory}
P.J. Flory, {\it Statistical mechanics of chain molecules},
Interscience, 1969.
  
\end{thebibliography}
\end{document}